\font\mybb=msbm10 at 12pt
\def\bbxx#1{\hbox{\mybb#1}}
\def\Z {\bbxx{Z}}
\def\R {\bbxx{R}}
\def\C {\bbxx{C}}

\def \aa {\alpha}
\def \bb {\beta}

\def \dd {\delta}
\def \ee {\epsilon}

\def \kk {\kappa}
\def \ll {\lambda}
\def \mm {\mu}
\def \nn {\nu}

\def \rr {\rho}
\def \ss {\sigma}

\def \th {\theta}

\def \ww{\omega}

 \def \ggg {\Gamma}
\def \ddd {\Delta}
\def \eee {\varepsilon}

\def \lll {\Lambda}
\def \wee {\wedge}
\def \sss {\Sigma}

\def \www{\Omega}

\def \ti {\tilde}

\def \2 {{1 \over 2}}
\def \3 {{1 \over 3}}
\def \4 {{1 \over 4}}
\def \5 {{1 \over 5}}
\def \6 {{1 \over 6}}
\def \7 {{1 \over 7}}
\def \8 {{1 \over 8}}
\def \9 {{1 \over 9}}
\def \0 { \infty}

\def\++ {{(+)}}
\def \- {{(-)}}
\def\+-{{(\pm)}}

\def\ek {\eqn\abc$$}

\def \pa {\partial}
\def\na {\nabla}


\tolerance=10000
\input phyzzx

 \def\unit{\hbox to 3.3pt{\hskip1.3pt \vrule height 7pt width .4pt \hskip.7pt
\vrule height 7.85pt width .4pt \kern-2.4pt
\hrulefill \kern-3pt
\raise 4pt\hbox{\char'40}}}

\def\nup#1({Nucl.\ Phys.\  {\bf B#1}\ (}

\REF\Sen {A. Sen, Nucl. Phys. {\bf B404} (1993) 109; Phys. Lett. {\bf 303B}
(1993); Int. J. Mod. Phys. {\bf A8} (1993) 5079; Mod. Phys. Lett. {\bf
A8} (1993) 2023;    Int. J. Mod. Phys. {\bf A9} (1994) 3707.}
\REF\HT{C.M. Hull and P.K. Townsend, Nucl. Phys. {\bf B438} (1995) 109.}
\REF\PKT{P.K. Townsend, Phys. Lett. {\bf B350}  (1995) 184.}
\REF\Witten{E. Witten, Nucl. Phys. {\bf B443} (1995)  85, hep-th/9503124.}
\REF\Democ{P.K. Townsend,  in the Proceedings of the
March 95 PASCOS/John Hopkins Conference,
hep-th/9507048.}
\REF\HUSC{C.M. Hull, in Proceedings of Strings 95.}
\REF\HStr{C.M. Hull, Nucl. Phys. {\bf B468} (1996) 113, hep-th/9512181.}
\REF\HTE{C.M. Hull and P.K. Townsend, Nucl. Phys. {\bf B451} (1995) 525,
hep-th/9505073.}
\REF\WitUSC{E. Witten, in Proceedings of Strings 95, hep-th/9507121.}
\REF\Oopen{C.M.Hull, Phys. Lett. {\bf B357} (1995) 545,
hep-th/9506194.}
\REF\dab{A. Dabholkar, Phys. Lett. {\bf B357} (1995) 307, hep-th/9506160.}
\REF\Sag{A. Sagnotti,  in "Non-Perturbative Quantum Field Theory", Proceedings
of
1987  Cargese Summer Institute, eds.
G. Mack et al  (Pergamon, 1988), p. 521.}
\REF\PW{J. Polchinski and E. Witten, Nucl. Phys. {\bf B460} (1996) 525. }
\REF\SGP {R. Sorkin, Phys. Rev. Lett. {\bf 51} (1983) 87 ; D. Gross and
M. Perry, Nucl. Phys. {\bf B226} (1983) 29.}
\REF\GH {G.W. Gibbons and C.M. Hull, Phys. Lett. {\bf 109B} (1982) 190.}
\REF\PPdual{E. Bergsoeff, I. Entrop and R. Kallosh, Phys. Rev.
{\bf D 49} (1994) 6663.}
\REF\TwAl{J. W. van Holten  and A. Van Proeyen    {\it Phys. A: Math. Gen.}
{\bf 15} (1982)
3763.}
\REF\DGHR {A. Dabholkar, G.W. Gibbons, J.A. Harvey and F. Ruiz-Ruiz,
Nucl. Phys. {\bf B340} (1990) 33.}
\REF\HS {G.T. Horowitz and A. Strominger, Nucl. Phys. {\bf B360} (1991)
197.}
\REF\CHS {C. Callan, J. Harvey and A. Strominger, Nucl. Phys. {\bf B359}
(1991)
611.}
\REF\DL {M.J. Duff and J.X. Lu, Nucl. Phys. {\bf B354} (1991) 141;
Phys. Rev. Lett. {\bf 66} (1991) 1402;
Phys. Lett. {\bf 273B} (1991) 409; M.J. Duff, G.W. Gibbons and P.K. Townsend,
Phys. Lett. {\bf 332B} (1994) 321.}
\REF\SFiv{
M.J. Duff, Class. Quant. Grav. {\bf 5} (1988) 189; A. Strominger, Nucl. Phys.
{\bf B343} (1990)
167; M.J. Duff and J.X. Lu, Nucl. Phys. {\bf B354} (1991) 129 and Nucl. Phys.
{\bf B357} (1991)
534. }
\REF\GGP{G.W. Gibbons, M.B. Green and M.J. Perry, Phys. Lett.
 {\bf B370} (1996) 37.}
\REF\Ebr{E. Bergshoeff, M. de Roo, M.B. Green, G. Papadopoulos and
P.K. Townsend,
Nucl. Phys.{\bf B470} {} {1996} 113.}
\REF\Witpq{E. Witten,
Nucl. Phys. {\bf B460} {} {1996} 335.}
\REF\Doug{M. Douglas and M. Li, hep-th/9604041.}
\REF\Ovr{B.A. Ovrut and D. Waldram, hep-th/9704045.}
\REF\Haw{S.W. Hawking, Phys. Lett. {\bf 60A} (1977) 81.}
\REF\AGH{C.M. Hull, Nucl. Phys. {\bf B260} (1985) 182 ;
L. Alvarez-Gaum\' e and P. Ginsparg, Commun. Math. Phys. {\bf 102} (1985) 311.}
\REF\RS{S. Ramaswamy and A. Sen, J. Math. Phys. {\bf 22} (1981) 2612.}
\REF\AS{A. Ashtekar and A. Sen, J. Math. Phys. {\bf 23} (1982) 2168.}
\REF\Nes{J.M. Nester, Phys. Lett. {\bf 83A}  (1981) 241.}
\REF\Hcmp{C.M. Hull, Commun. Math. Phys. {\bf 90}  (1983) 545.}
\REF\GHT{Gibbons G W, Horowitz G T and Townsend P K, {\it Class. Quantum Grav.}
{\bf 12} (1995) 297.}
\REF\Witpos{E.Witten, Commun. Math. Phys. {\bf 80}  (1981) 381.}
\REF\Wex{T. Parker and C. Taubes, Commun. Math. Phys. {\bf 84} (1982) 223; O.
Reula, J. Math. Phys. {\bf 23}(1982) 5.}
\REF\GP {G.W. Gibbons and M.J. Perry, Nucl. Phys. {\bf B248} (1984) 629.}
\REF\LS{J. Lee and R.D. Sorkin, Commun. Math. Phys. {\bf 116} (1988) 353.}
\REF\MSp{O.M. Moreschi and G.A.J. Sparling, J. Math. Phs. {\bf 27} (1986)
2402.}
\REF\Bom{L. Bombelli, R.K. Koul, G. Kunstatter, J. Lee and R.D. Sorkin, Nucl.
Phys. {\bf B289} (1987) 735.}
\REF\BM{B. Jensen and P.V. Mani, hep-th/9702044.}
\REF\Ort{R. Kallosh, D. Kastor, T. Ortin and T. Torma, Phys. Rev. {\bf D50}
(1944) 6374.}
\REF\WitKK{E. Witten, Nucl. Phys. {\bf B289} (1982) 735.}
\REF\MTR{M.M. Taylor-Robinson, hep-th/9609234.}
\REF\GHI{G.W. Gibbons and S.W. Hawking, Commun. Math. Phys. {\bf 61} (1979)
291.}
\REF\Se{A.Sen, Nucl. Phys. {\bf B474} (1996) 361, hep-th/9604070; Nucl. Phys.
{\bf B475} (1996)
562,  hep-th/9605150; hep-th/9609176}
\REF\Twe{M. Blencowe, M.J. Duff, C.M. Hull and K.S. Stelle, unpublished; M.
Blencowe and M.J. Duff,
Nucl. Phys. {\bf 310} (1988) 387;    D. Kutasov and E.
Martinec, hep-th/9510182;  D.\ Kutasov, E.\ Martinec and M.\ O'Loughlin,
hep-th/9603116; I. Bars,
hep-th/9604200.}
\REF\Tse{A.A. Tseytlin,  Nucl.
Phys. {\bf B469} (1996) 51.}
\REF\Rom{L. Romans, Phys. Lett. {\bf 169B} (1986) 374.}
\REF\Dua{
 M.\ Aganagic, J.\ Park, C.\ Popescu and J.\ Schwarz,  hep-th/9702133; M. Abou
Zeid and C.M. Hull,
hep-th/9704021; M.B. Green and M. Gutperle, Phys. Lett. {\bf B377} (1996) 28;
Y. Lozano,
hep-th/9701186.}
\REF\Rub{P. Ruback, Commun. Math. Phys. {\bf 107} (1986) 93.}
 \REF\Hwave{C.M. Hull, Phys. Lett. {\bf B139} (1984) 39.}
\REF\BHO{E. Bergsoeff, C.M. Hull and T. Ortin, Nucl. Phys. {\bf 451} (1995)
547.}
\REF\BR{E. Bergshoeff, M. de Roo, E. Eyras, B. Janssen and J.P. van de Schaar,
hep-th/9704120.}
\REF\Berg{E. Bergshoeff, hep-th/9607238, 9611099.}
\REF\Strom{A. Strominger, Phys. Lett. {\bf B383} (1996) 44.}
\REF\GHaw{G.W. Gibbons and S.W. Hawking, Phys. Lett. {\bf 78B} (1978) 430.}
\REF\BSV{M. Berschadsky, V. Sadov and C. Vafa, Nucl. Phys.
{\bf B463} (1996) 398, hep-th/9510225.}
\REF\Des{K. Bautier, S. Deser, M. Henneaux and D. Seminara, hep-th/9704131.}
\REF\PKTlecs{P.K. Townsend, lectures at the Euroconference: Duality and
Supersymmetric Theories, Newton Institute, 1997.}
\REF\Winstanton{E. Witten, Nucl. Phys. {\bf B460} (1996) 541.}


\Pubnum{ \vbox{ \hbox {QMW-97-16}  \hbox {NI97028-NQF}\hbox{hep-th/9705162}} }
\pubtype{}
\date{May, 1997}

\titlepage

\title {\bf  Gravitational Duality, Branes and Charges}

\author{C.M. Hull}
\address{Physics Department,
Queen Mary and Westfield College,
\break
Mile End Road, London E1 4NS, U.K.}
\andaddress{Isaac Newton Institute, 20 Clarkson Road,
\break
Cambridge CB3 0EH, U.K.}
\vskip 0.5cm

\abstract {

It is argued that $D=10$ type II strings and  M-theory in $D=11$ have   $D-5$
branes and 9-branes
that are not standard $p$-branes coupled to anti-symmetric tensors.
The global charges in a $D$-dimensional theory of gravity consist of a momentum
$P_M$ and a dual
$D-5$ form charge
$K_{M_1...M_{D-5}}$, which is related to  the NUT charge.
On dimensional reduction, $P$ gives the electric charge and $K$ the
  magnetic charge of the graviphoton.
The charge $K$ is constructed and shown to occur in the superalgebra and BPS
bounds in $D\ge 5$,
and leads to a NUT-charge modification of the BPS bound in $D=4$.
$K$ is carried by Kaluza-Klein monopoles, which can be regarded as $D-5$
branes.
Supersymmetry and U-duality imply that the type IIB theory has $(p,q)$
9-branes.
Orientifolding  with 32 (0,1) 9-branes gives the type I string, while modding
out by a
 related discrete symmetry with 32 (1,0) 9-branes gives the $SO(32)$
heterotic string. Symmetry enhancement, the effective world-volume theories and
the possibility
of a twelve dimensional origin are discussed. }

\endpage

%
%
%
%
%
\newhelp\stablestylehelp{You must choose a style between 0 and 3.}%
\newhelp\stablelinehelp{You should not use special hrules when stretching
a table.}%
\newhelp\stablesmultiplehelp{You have tried to place an S-Table inside another
S-Table.  I would recommend not going on.}%
%
%
\newdimen\stablesthinline
\stablesthinline=0.4pt
\newdimen\stablesthickline
\stablesthickline=1pt
%
%
\newif\ifstablesborderthin
\stablesborderthinfalse
\newif\ifstablesinternalthin
\stablesinternalthintrue
\newif\ifstablesomit
\newif\ifstablemode
\newif\ifstablesright
\stablesrightfalse
%
%
\newdimen\stablesbaselineskip
\newdimen\stableslineskip
\newdimen\stableslineskiplimit
%
%
\newcount\stablesmode
\newcount\stableslines
\newcount\stablestemp
\stablestemp=3
\newcount\stablescount
\stablescount=0
\newcount\stableslinet
\stableslinet=0
%
%
%
\newcount\stablestyle
\stablestyle=0
%
%
\def\stablesleft{\quad\hfil}%
\def\stablesright{\hfil\quad}%
%
%
\catcode`\|=\active%
%
%
\newcount\stablestrutsize
\newbox\stablestrutbox
\setbox\stablestrutbox=\hbox{\vrule height10pt depth5pt width0pt}
\def\stablestrut{\relax\ifmmode%
                         \copy\stablestrutbox%
                       \else%
                         \unhcopy\stablestrutbox%
                       \fi}%
%
%
\newdimen\stablesborderwidth
\newdimen\stablesinternalwidth
\newdimen\stablesdummy
\newcount\stablesdummyc
\newif\ifstablesin
\stablesinfalse
%
%
\def\begintable{\stablestart%
  \stablemodetrue%
  \stablesadj%
  \halign%
  \stablesdef}%
\def\stablesadj{%
  \ifcase\stablestyle%
    \hbox to \hsize\bgroup\hss\vbox\bgroup%
  \or%
    \hbox to \hsize\bgroup\vbox\bgroup%
  \or%
    \hbox to \hsize\bgroup\hss\vbox\bgroup%
  \or%
    \hbox\bgroup\vbox\bgroup%
  \else%
    \errhelp=\stablestylehelp%
    \errmessage{Invalid style selected, using default}%
    \hbox to \hsize\bgroup\hss\vbox\bgroup%
  \fi}%
\def\stablesend{\egroup%
  \ifcase\stablestyle%
    \hss\egroup%
  \or%
    \hss\egroup%
  \or%
    \egroup%
  \or%
    \egroup%
  \else%
    \hss\egroup%
  \fi}%
\def\stablestart{%
  \ifstablesin%
    \errhelp=\stablesmultiplehelp%
    \errmessage{An S-Table cannot be placed within an S-Table!}%
  \fi
  \global\stablesintrue%
  \global\advance\stablescount by 1%
  \message{<S-Tables Generating Table \number\stablescount}%
  \begingroup%
  \stablestrutsize=\ht\stablestrutbox%
  \advance\stablestrutsize by \dp\stablestrutbox%
  \ifstablesborderthin%
    \stablesborderwidth=\stablesthinline%
  \else%
    \stablesborderwidth=\stablesthickline%
  \fi%
  \ifstablesinternalthin%
    \stablesinternalwidth=\stablesthinline%
  \else%
    \stablesinternalwidth=\stablesthickline%
  \fi%
  \tabskip=0pt%
  \stablesbaselineskip=\baselineskip%
  \stableslineskip=\lineskip%
  \stableslineskiplimit=\lineskiplimit%
  \offinterlineskip%
  \def\borderrule{\vrule width \stablesborderwidth}%
  \def\internalrule{\vrule width \stablesinternalwidth}%
  \def\thinline{\noalign{\hrule height \stablesthinline}}%
  \def\thickline{\noalign{\hrule height \stablesthickline}}%
  \def\trule{\omit\leaders\hrule height \stablesthinline\hfill}%
  \def\ttrule{\omit\leaders\hrule height \stablesthickline\hfill}%
  \def\tttrule##1{\omit\leaders\hrule height ##1\hfill}%
  \def\stablesel{&\omit\global\stablesmode=0%
    \global\advance\stableslines by 1\borderrule\hfil\cr}%
  \def\el{\stablesel&}%
  \def\elt{\stablesel\thinline&}%
  \def\eltt{\stablesel\thickline&}%
  \def\elttt##1{\stablesel\noalign{\hrule height ##1}&}%
  \def\elspec{&\omit\hfil\borderrule\cr\omit\borderrule&%
              \ifstablemode%
              \else%
                \errhelp=\stablelinehelp%
                \errmessage{Special ruling will not display properly}%
              \fi}%
  \def\stmultispan##1{\mscount=##1 \loop\ifnum\mscount>3 \stspan\repeat}%
  \def\stspan{\span\omit \advance\mscount by -1}%
  \def\multicolumn##1{\omit\multiply\stablestemp by ##1%
     \stmultispan{\stablestemp}%
     \advance\stablesmode by ##1%
     \advance\stablesmode by -1%
     \stablestemp=3}%
  \def\multirow##1{\stablesdummyc=##1\parindent=0pt\setbox0\hbox\bgroup%
    \aftergroup\emultirow\let\temp=}
  \def\emultirow{\setbox1\vbox to\stablesdummyc\stablestrutsize%
    {\hsize\wd0\vfil\box0\vfil}%
    \ht1=\ht\stablestrutbox%
    \dp1=\dp\stablestrutbox%
    \box1}%
  \def\stpar##1{\vtop\bgroup\hsize ##1%
     \baselineskip=\stablesbaselineskip%
     \lineskip=\stableslineskip%
     \lineskiplimit=\stableslineskiplimit\bgroup\aftergroup\estpar\let\temp=}%
  \def\estpar{\vskip 6pt\egroup}%
  \def\stparrow##1##2{\stablesdummy=##2%
     \setbox0=\vtop to ##1\stablestrutsize\bgroup%
     \hsize\stablesdummy%
     \baselineskip=\stablesbaselineskip%
     \lineskip=\stableslineskip%
     \lineskiplimit=\stableslineskiplimit%
     \bgroup\vfil\aftergroup\estparrow%
     \let\temp=}%
  \def\estparrow{\vfil\egroup%
     \ht0=\ht\stablestrutbox%
     \dp0=\dp\stablestrutbox%
     \wd0=\stablesdummy%
     \box0}%
  \def|{\global\advance\stablesmode by 1&&&}%
  \def\|{\global\advance\stablesmode by 1&\omit\vrule width 0pt%
         \hfil&&}%
  \def\vt{\global\advance\stablesmode by 1&\omit\vrule width \stablesthinline%
          \hfil&&}%
  \def\vtt{\global\advance\stablesmode by 1&\omit\vrule width
\stablesthickline%
          \hfil&&}%
  \def\vttt##1{\global\advance\stablesmode by 1&\omit\vrule width ##1%
          \hfil&&}%
  \def\vtr{\global\advance\stablesmode by 1&\omit\hfil\vrule width%
           \stablesthinline&&}%
  \def\vttr{\global\advance\stablesmode by 1&\omit\hfil\vrule width%
            \stablesthickline&&}%
  \def\vtttr##1{\global\advance\stablesmode by 1&\omit\hfil\vrule width ##1&&}%
  \stableslines=0%
  \stablesomitfalse}
\def\stablesdef{\bgroup\stablestrut\borderrule##\tabskip=0pt plus 1fil%
  &\stablesleft##\stablesright%
  &##\ifstablesright\hfill\fi\internalrule\ifstablesright\else\hfill\fi%
  \tabskip 0pt&&##\hfil\tabskip=0pt plus 1fil%
  &\stablesleft##\stablesright%
  &##\ifstablesright\hfill\fi\internalrule\ifstablesright\else\hfill\fi%
  \tabskip=0pt\cr%
  \ifstablesborderthin%
    \thinline%
  \else%
    \thickline%
  \fi&%
}%
\def\endtable{\advance\stableslines by 1\advance\stablesmode by 1%
   \message{- Rows: \number\stableslines, Columns:  \number\stablesmode>}%
   \stablesel%
   \ifstablesborderthin%
     \thinline%
   \else%
     \thickline%
   \fi%
   \egroup\stablesend%
\endgroup%
\global\stablesinfalse}
%
%

\chapter {Introduction}

BPS states have played a vital role in the unravelling of the non-perturbative
structure of
superstring theory and $M$-theory [\Sen-\PW]. As these break some fraction of
the supersymmetry, they
carry some charge that appears on the right hand side of the global
supersymmetry algebra.
The standard example  of such a charge is the
 electric
or magnetic charge  for some $n$-form anti-symmetric tensor gauge field. This
can be a central
charge, as in the case of the electric charge for a vector gauge field, but in
general it is a
$p$-form charge, and the standard example of a BPS state carrying a $p$-form
charge is a $p$-brane
-- an extended  object whose world-volume is a $p+1$   dimensional space of
Lorentzian signature.
For an $n$-form anti-symmetric tensor gauge field in $D$ dimensions, the
electric charge is carried
by a $p$-brane with $p=n-1$, while the magnetic charge is is carried
by a $\ti p$-brane with $\ti p=D-n-3$.

The purpose of this paper is to re-examine the spectrum of BPS branes in string
theory and
M-theory. In section 2,  the charges that occur in the corresponding
supersymmetry
algebras are studied. In addition to the expected $p$-brane charges, there is
an extra $D-5$ form
charge
$K$ in
$D$ dimensions, and an extra 9-form charge for the type IIB theory and for type
IIA or M-theory.

The $D-5$ form charge is carried by Kaluza-Klein (KK) monopole space-times
[\SGP]. For example, the
Kaluza-Klein monopole space-time $N_4\times \R^{D-5,1}$, where $N_4$ is
Euclidean Taub-NUT space, can
be thought of as a $D-5$ brane and carries a $D-5$ form charge proportional to
the volume form of
$\R^{D-5}$.
On dimensional reduction to $D-1$ dimensions, the KK monopole gives rise to
magnetic charge of the
graviphoton, which in $D-1$ dimensions is a $(D-1)-4$ form charge $Z_{D-5}$.
On reduction, the $D-5$ form $K_{D-5}$ gives the magnetic charge $Z_{D-5}$ and
the $K$-charge
for $D-1$ dimensions, which is a $(D-1)-5$ form charge.

Thus in $D$ dimensions, gravity has two global charges, the ADM momentum $P_M$
and the new charge
$K_{M_1...M_{D-5}}$ which are \lq dual' in the same way that electric and
magnetic charges
are dual in Maxwell theory. Indeed, on dimensional reduction, $P_M$
gives the electric charges
and
$K_{M_1...M_{D-5}}$ gives the magnetic charges of the graviphotons.
Kaluza-Klein monopole space-times $N_4\times \R^{D-5,1}$
can be thought of as  $D-5$ branes occurring in the purely gravitational sector
of the theory; this
was suggested in [\HT], where it was shown that the    spectrum of compactified
string
theories included BPS states arising from wrapping such $D-5$ branes.  In
section 3, the charge
$K_{M_1...M_{D-5}}$ is constructed explicitly and
is
shown to occur  in   superalgebras evaluated in
suitable backgrounds. The K-charge is related to the NUT charge, and the
supersymmetry-motivated
formula for $K$ gives a new expression for the NUT charge, which can be used to
prove the
generalisation of the gravitational Bolgomolnyi bound [\GH] due to NUT charge;
this is done in
section 4.
 In section
5, the charge $K$ is calculated for certain examples.
The  interpretation of $N_4\times \R^{D-5,1}$ as a $D-5$ brane would suggest
that its collective
coordinates should form a $D-4$ dimensional supermultiplet including $4$
scalars, corresponding to
the position of the brane in $N_4$. However, no such multiplet exists in e.g. 7
dimensions
(corresponding to $D=11$). This problem is addressed and resolved in section 8,
where the
collective coordinate structure  is found. There is a sense in which the KK
monopole
can be regarded as a \lq twisted' $D-4$ brane  [\HT], and the number of
translational zero modes is
that appropriate for a $D-4$ brane.

The configurations that preserve half the supersymmetry include pp-wave
configurations in addition
to the   $p$-branes and KK monopoles. On   compactification, the set of all BPS
states in
the  compactified theory includes   $p$-branes and KK monopoles wrapped around
internal dimensions,
together with pp-waves moving in the compactifying space [\HT]. These are all
on an equal footing
in the compactified theory, and related by U-duality [\HT]; in particular,
pp-waves in an
internal direction are related to fundamental strings by T-duality [\PPdual].
There is then a
sense in which pp-waves are a type of string, so that pp-waves and KK monopoles
can be thought of
as gravitational branes, or \lq G-branes'.  The pp-waves carry the charge $P$
and KK monopoles carry
the dual charge
$K$, and on compactification these G-branes are U-dual to
$p$-branes which couple to anti-symmetric tensor gauge fields [\HT].

The type IIB theory has $p$-branes for $p=1,3,5,7,9$ and in particular has a
Dirichlet 9-brane,
which gives rise to Chan-Paton factors for open strings. The IIB superalgebra
implies
that the $p$-form charges carried by $p$-branes are singlets under the
$SL(2,\Z)$
duality symmetry for $p=3,7$ and are doublets for $p=1,5,9$.
For $p=1,5$, this is in accord with the existence of $(p,q)$ strings and
$(p,q)$
5-branes [\Witpq], with charge $p$ with respect to the NS-NS (Neveu-Schwarz
Neveu-Schwarz) 2-form and
charge
$q$
 with respect to the RR (Ramond-Ramond) 2-form.
The algebra suggests that there should also be $(p,q)$ 9-branes.
Indeed, acting on the Dirichlet
9-brane with $SL(2,\Z)$ must either leave the brane invariant, or give a new
9-brane,
and the superalgebra implies that the 9-brane charge must transform, so that
the action
of $SL(2,\Z)$ on the Dirichlet 9-brane generates $(p,q)$ 9-branes.

Of course, this is rather formal as it is not consistent to have Dirichlet
9-branes in general
 backgrounds.
However, orientifolding the type IIB string to give the type I string [\Sag],
i.e. modding out the  by the
perturbative $\Z_2$  world-sheet parity reversal symmetry   $\www$, requires
the presence of 32
Dirichlet 9-branes, which give rise to   $SO(32)$ Chan-Paton factors [\PW].
This suggests that the new 9-branes might   play a role in similar
constructions. This is addressed
in section 6, where it is shown that
 there must be a non-perturbative $\Z_2$ symmetry
$\ti \www$ such that modding out by $\ti \www$ in the presence of 32
$(1,0)$ 9-branes
gives the $SO(32)$ heterotic string.
The strong-weak coupling duality of type IIB interchanges the $(0,1)$ 9-branes
with the $(1,0)$
9-branes and $\www$ with $\ti \www$, so that this is consistent with
 the duality between type I and $SO(32)$ heterotic strings
[\Witten,\Oopen-\PW].
The new $(1,0)$ 9-branes are essential for the construction and there are
analogous constructions
using $(p,q)$ 9-branes.

In section 7, it is argued using T-duality that the presence of a new (1,0)
9-brane in type IIB
theory implies the existence of a new 9-brane in the IIA theory, which should
originate
from a 9-brane in M-theory; in section 2, it is shown that such 9-form charges
can indeed occur in
the relevant superalgebras. The possibility of a 9-brane in M-theory has also
been discussed in [\Ebr,\BR].
In section 8, the effective world-volume theories describing the collective
coordinates for all
branes, including  G-branes, are found, with supersymmetry implying that the KK
monopole has less
translational zero-modes than might have been expected naively.
 All M-theory charges could naturally arise from  a 2-form, a self-dual
6-form and  a 10-form charge  in 12 dimensions, but  there is no way the
effective
world-volume theories of all M-branes can arise from covariant dynamics of
branes in twelve
dimensions. In section 9, the gauge symmetry enhancement from coincident
G-branes is discussed.

With these \lq new branes' included, all the charges that can occur on the
right hand side of the
superalgebra in fact do occur, and are needed to get the right counting on
compactification, where
the new branes give rise to branes related to known ones by U-duality.

\chapter {Superalgebras and Brane Charges}

The most general super-Poincar\' e algebra  in $D$ dimensions
includes some anti-symmetric tensor charges in the anti-commutator of two
supercharges, and these commute with all generators except the Lorentz
generators.
The anti-commutator  of two supercharges  gives\foot{The notation and
conventions are as in [\Democ].
The vector indices are $M=0,\dots D-1$ and the spinor indices are $\aa, \bb$.
The charge conjugation
matrix is
$C_{\aa \bb}$ and
$\Gamma^{MN\dots P}=\Gamma^{[M}\Gamma^N\dots  \Gamma^{P]}$.}
 a symmetric bi-spinor $Z_{\aa\bb}$ defined by
$\{Q_\alpha ,Q_\beta \}=Z_{\aa\bb}$, and $Z_{\aa\bb}$ can be rewritten as
a sum of terms, each of which is of the form
$Z_{MN...Q}\big(\Gamma^{MN...Q}C\big)_{\alpha\beta}$
for some set of $p$-form charges $Z_{M_1...M_p}$. The values of $p$ which occur
will depend on the
dimension considered, but for a conventional superalgebra will include a 1-form
corresponding to
the space-time momentum $P_M$.
For a theory with $N$ supercharges $Q_\aa^a$ with $a=1,...,N$, the
extended superalgebra
includes the anti-commutator
 $\{Q_\alpha ^a,Q_\beta ^b\}=Z_{\aa\bb}^{ab}+\bar Z_{\aa\bb}^{ab}$
 where $Z_{\aa\bb}^{ab}=Z_{\aa\bb}^{ba}=Z_{\bb\aa}^{ab}$
and $\bar Z_{\aa\bb}^{ab}=-\bar Z_{\aa\bb}^{ba}=-\bar Z_{\bb\aa}^{ab}$.

 In $D=11$ the $N=1$ superalgebra has one 32-component
Majorana supercharge
$Q_\aa$ and these satisfy the anti-commutation relations [\TwAl,\Democ]
$$
\{Q_\alpha,Q_\beta\} = \big(\Gamma^MC\big)_{\alpha\beta}P_M +
{1\over  2 !}
\big(\Gamma^{MN}C\big)_{\alpha\beta}\, Z_{MN} +{1\over  5 !}
\big(\Gamma^{MNPQR}C\big)_{\alpha\beta}\, Z_{MNPQR}\ .
\eqn\el$$
The left-hand side is   symmetric   in its spinor indices and so  has
$32.33/2=528$ components
in general. Anti-symmetric tensor charges are included on the right-hand-side
to give a total of
$11 + 55 + 462 = 528$ components.
Dimensionally reducing to $D=10$ gives the IIA algebra, with the
anti-commutator of two Majorana
supercharges being
$$
\eqalign{
\{Q_\alpha,Q_\beta\} =& \big(\Gamma^MC\big)_{\alpha\beta} P_M
+ (\Gamma_{11}C)_{\alpha\beta}Z + \big(\Gamma^M\Gamma_{11}C\big)_{\alpha\beta}
Z_M +{1\over   2!} \big(\Gamma^{MN}C\big)_{\alpha\beta}Z_{MN}
\cr &+\,{1\over   4!} \big(\Gamma^{MNPQ}\Gamma_{11}C\big)_{\alpha\beta}Z_{MNPQ}
+ {1\over   5!}\big(\Gamma^{MNPQR}C\big)_{\alpha\beta}Z_{MNPQR}\ .
\cr}
\eqn\toa
$$
 The IIB supersymmetry algebra has two Majorana-Weyl supercharges,
$Q_\alpha^a$, (a=1,2), of the same chirality with
$$\eqalign{
\{Q_\alpha^a,Q_\beta^b\} &=\, \delta^{ab}\big({\cal
P}\Gamma^MC\big)_{\alpha\beta} P_M + \big({\cal
P}\Gamma^MC\big)_{\alpha\beta}\,
  Z_M^{ab} + {1\over   3!}\varepsilon^{ab}\big({\cal
P}\Gamma^{MNP}C\big)_{\alpha\beta}\, Z_{MNP}
\cr
& +{1\over   5!}\delta^{ab} \big({\cal
P}\Gamma^{MNPQR}C\big)_{\alpha\beta}(Z^+)_{MNPQR}
\cr &
+
{1\over   5!}\big({\cal P}\Gamma^{MNPQR}C\big)_{\alpha\beta}(
Z^+)_{MNPQR}^{ab}\;
\cr}
\eqn\tob$$
where $Z_M^{ab}$ and $( Z^+)_{MNPQR}^{ab}$ are $2\times 2$ traceless symmetric
matrices in $a,b$,
$\dd_{ab}Z_M^{ab}=0$ and $\dd_{ab}( Z^+)_{MNPQR}^{ab}=0$,
 and ${\cal P}= \2 (1+\ggg ^{11})$ is the chiral projector.
All three of the 5-form charges $(Z^+)_{MNPQR},(Z^+)_{MNPQR}^{ab}$ are
self-dual 5-forms,
$*Z^+=Z^+$. The total number of components of all
charges on the RHS of \tob\  is
$
10 + 2\times 10 + 120 + 126 + 2\times 126 = 528 $, which again balances the
number of components on
the LHS.

The $P_M$ on the RHS of \el-\tob\ is the $D$-momentum while some of the
$p$-form charges
$Z_{M_1\dots M_p}$   are carried by $p$-branes.
An $n+1$-form   field strength $F_{M_1\dots M_{n+1}}$ satisfies  the field
equations
$$\na _N (* \ti F)^{N{M_1\dots M_n}}=J^{M_1\dots M_n}
\eqn\abc$$
and
$$\na _N (* F)^{N{M_1\dots M_{\ti n }}} =\ti J^{M_1\dots M_{\ti n}}
\eqn\abc$$
where $\ti n=D-n-2$,
and $\ti F$ is given by varying the action $S$ with respect to $F$:
$$
\ti F^{ M_1\dots M_
{\ti n+1}
 }\equiv
{1 \over (n+1)!}
\varepsilon ^{ M_1\dots M_
{\ti n+1}   N_1\dots N_
{  n+1}
}
{\dd S \over \dd F_
{ N_1\dots
N_{  n+1}}
}= *  F^{ M_1\dots M_{\ti n+1}}+\dots
\eqn\abc$$
so that  $\ti F = *F+ \dots$ where $*F $ is the Hodge
dual, and the dots denote terms dependent on the interactions;    for the free
 theory,  $\ti F =
*F$. The currents $J,\ti J$
  satisfy the conservation laws
$$
\na_{M_1}J^{M_1\dots M_n}=0, \qquad
\na_{M_1}
\ti J^{M_1\dots M_{\ti n}}=0
\eqn\abc$$
In regions in which the magnetic source $\ti J $ vanishes, $F$ can be written
in terms of an
$n$-form potential
$A$,  $F=dA$, while in regions in  which the electric source $ J $ vanishes,
$\ti F$ can be written
in terms of a
$\ti n$-form potential $\ti A$, $\ti F=d\ti A$.

An electric $p$-brane source couples to a $n=p+1$ form potential $A$ through
the Wess-Zumino
term\foot{The alternating symbol $\ee ^{ M_1\dots M_
{\ti n+1}}$ is a tensor density while
$\varepsilon ^{ M_1\dots M_
{\ti n+1}}$ is a tensor.}
$$\int d^n \ss \, A_{M_1 \dots M_n}\ee ^{m_1 \dots m_n} \pa _{m_1} X^{M_1}\dots
\pa _{m_n} X^{M_n}
\eqn\abc$$
where $\ss^m$ are world-volume coordinates, $m=1, \dots , n=p+1$, and
$X^M(\ss)$ is the position of
the brane.
This then leads to an $n$-form current density
$$J^{M_1 \dots M_n}(X)
=\int d^n \ss \, \ee ^{m_1 \dots m_n} \pa _{m_1} X^{M_1}\dots \pa _{m_n}
X^{M_n} \dd (X^M- X^M(\ss))
\eqn\abc$$
A magnetic $p=\ti n -1$ brane source couples to $\ti A$ and leads to a similar
formula for $\ti J$,
with $n$
replaced by
$\ti n$.
A static  configuration carries an electric charge density $z_{i_1\dots
i_{p}}=J_{0i_1\dots i_{p}}$
and a magnetic charge density
$\ti z_{i_1\dots i_{\ti p}}=\ti J_{0i_1\dots i_{\ti p}}$
where $n=p+1$, $\ti n = \ti p +1$ and $i,j= 1, \dots D-1$ are spatial indices.
These charge densities give rise to total charges
$Z_{i_1\dots i_{p}}$,
$\ti Z_{i_1\dots i_{\ti p}}$ which are again anti-symmetric tensors carrying
purely spatial indices.

These $\dd $-function sources can often be replaced by solitonic $p$-brane
solutions of the
theory [\DGHR-\Ebr].
A spatial slice of a $p$-brane configuration has an asymptotic \lq boundary' at
spatial infinity
given by $\R ^p \times S ^{D-p-2}$ for an infinite brane, or by
 $\R^{p-q}\times T ^{q} \times S ^{D-p-2}$ if the $p$-brane is wrapped around
$q$ toroidal dimensions.
The  charge of an electric $p$-brane coupled to a
$n=p+1$ form potential $A$ and
aligned in the directions given by the
spatial
$p$-form   $v$ is given by
$$
Z.v \equiv {1\over p!}Z_{i_1\dots i_{p}} v^{i_1\dots i_{p}}
 = {1 \over \www_{\ti n +1}}
\int _{S^{\ti n+1}}\ti F
\eqn\abc$$
where the integral is over the $\ti n+1$ sphere at transverse spatial infinity
surrounding
the brane and $\www_d$ is the area of the unit $d$-sphere. Similarly, the
charge
of a  magnetic $p$-brane coupled to an
$\ti n$-form potential $A$ with $\ti n=D-n-2=p+1$
 is
$$
\ti Z.v\equiv {1\over p!}\ti Z_{i_1\dots i_{p}} v^{i_1\dots i_{p}}={1 \over
\www_{  n +1}} \int _{S^{
n+1}} F
\eqn\abc$$
where the integral is over the $  n+1$ sphere at transverse spatial infinity
surrounding the brane.

A $p$-brane carries a $p$-form charge $Z_{i_1...i_p}$ with purely spatial
indices.
The spatial components of the  $Z_{M_1..M_p}$ appearing in the algebras
\el-\tob\ can arise as
$p$-brane charges,
but this leaves the question of the interpretation of the
components $Z_{0i_1...i_q}$.
These can
be dualised to   a spatial $\hat q$-form
charge
$\hat Z_{i_1...i_{\hat q}}$ given  by the spatial components of $*   Z$,
$$
\hat Z_{j_1...j_{\hat q}}={1 \over  (q +1 )!}\eee _{{j_1...j_{\hat
q}}0i_1...i_q  }
Z^{0i_1...i_q}
\eqn\abc$$
 which could arise as a $\hat q= D-q-1$ brane
charge, and this is the most natural interpretation of the charge
$Z_{0i_1...i_q}$.\foot{This
interpretation has been suggested independently by Paul Townsend.} We can now
compare this with the
expected
$p$-brane spectrum for M-theory and type II strings.

Consider first $D=11$; M-theory has a membrane and a 5-brane coupling to the
3-form gauge field. The
algebra
\el\ has a 2-form $Z_{MN}$ charge
which gives a  2-brane charge  $Z_{ij}$ and a 9-brane  charge
$\hat Z_{i_1...i_9}$,
and a 5-form charge
$Z_{MNPQR}$, which gives  a 5-brane charge
$Z_{ijklm}$   and a 6-brane charge $\hat Z_{i_1...i_6}$.
Thus the superalgebra has charges that could correspond to an extra 6-brane and
an extra 9-brane,
so that the question arises as to whether these actually occur and, if so,
what is their interpretation.

For the type IIA theory, the expected branes are a string and a 5-brane,
together with D-branes for
$p=0,2,4,6,8$.
The central charge $Z$ in \toa\ is the $0$-brane charge, the 4-form charge
$Z_{MNPQ}$ gives a 4-brane
charge
$Z_{ijkl}$ and a 6-brane charge
$\hat Z_{ijklmn}$, while the 2-form charge $Z_{MN }$ gives a 2-brane
charge
$Z_{ij }$ and an 8-brane charge
$\hat Z_{ijklmnpq}$.
The 5-form charge
  $Z_{MNPQR}$ gives {\it two} 5-brane
charges,
$Z_{ijklm}$ and
$\hat Z_{ijklm}$, while
the 1-form charge $Z_{M}$ gives a string
charge
$Z_{i }$ and a 9-brane charge
$\hat Z_{i_1...i_9}$. Comparing with the expected brane scan, we are finding
charges that
could correspond to
an extra
5-brane and an extra 9-brane.

Finally, for the type IIB string, the expected branes are a string and a
5-brane, together with
D-branes for
$p=1,3,5,7,9$ carrying Ramond-Ramond charge. There are also $(p,q)$ strings
carrying NS-NS string
charge $p$
and RR string charge $q$. In the quantum theory, $q,p$ are integers and
non-trivial bound states
occur whenever $p$ and $q$ are co-prime [\Witpq]. Similarly, there are $(p,q)$
5-branes, and for both
strings and 5-branes the 2-vector of charges $(p,q)$ transforms as a doublet
under the action of
$SL(2,\Z)$ U-duality transformations.
In the algebra \tob, the the 3-form charge $Z_{MNP}$ gives a 3-brane
charge
$Z_{ijk}$ and a 7-brane charge
$\hat Z_{ijklmnp}$. The two 1-form charges $  Z_M^{ab}$ give two string charges
$  Z_i^{ab}$
 and   two 9-brane charges
$\hat Z_{i_1...i_9}^{ab}$.
A general 5-form charge $Z_{MNPQR}$ gives {\it two} 5-brane charges,
$Z_{ijklm}$ and
$\hat Z_{ijklm}$, but for a self-dual 5-form charge, these are equal,
$Z_{ijklm}=\hat Z_{ijklm}$.
Thus the three self-dual  5-form charges
$(Z^+)_{MNPQR},(Z^+)_{MNPQR}^{ab}$   give the charges for 3 five-branes,
$(Z^+)_{ijklm},(Z^+)_{ijklm}^{ab}$.

The $SL(2,\R)$ symmetry of the supergravity equations of motion acts through a
compensating $SO(2)$
transformation in symmetric gauge, so that the index $a=1,2$ on the
supercharges
undergoes $SO(2)$ rotations. The two charges $  Z_i^{ab}$
are a doublet of $SO(2)$ and arise from charges which are a doublet of $SL(2)$
(see section 6), and
correspond to the charges of
$(p,q)$ strings. Similarly, the two charges   $(Z^+)_{ijklm}^{ab}$ correspond
to the charges of
$(p,q)$ 5-branes.  Similarly, the two 9-brane charges
$\hat Z_{i_1...i_9}^{ab}$ could correspond to the charges of $(p,q)$ 9-branes.
Note that the 3-brane and 7-brane charges are singlets under the action of
$SO(2)$.
This was expected for the 3-brane since the 4-form gauge field to which it
couples does not
transform under $SL(2,\R)$. The 7-brane couples to the axion which does
transform under
$SL(2,\R)$. Acting on the 7-brane solutions of [\GGP] with $SL(2,\Z)$
gives a  new 7-brane with the same 7-brane charge (the coupling to the 8-form
potential which is
dual to the axion is unchanged) but the $SL(2,\Z)$ monodromy changes and the
couplings to other
branes also changes [\Doug].
In the IIB algebra, there are thus charges for an extra 5-brane and for $(p,q)$
9-branes.

We now turn to the interpretation of the extra charges that arise in the above
cases.
We have seen that in $D=10,11$ dimensions, there is an extra spatial
$(D-5)$-form charge
that does not seem to have an interpretation in terms of
electric or magnetic $p$-brane charges -- we have already accounted for all of
these.
In  the following sections, it will be seen that this  $(D-5)$-form charge is
non-zero for
 Kaluza Klein
monopole space-times in which all the anti-symmetric tensor gauge fields
vanish.
It will be argued that in any theory of gravity in $D>4$ dimensions, there is,
in addition to the
ADM momentum
$P_M$, a $(D-5)$-form charge $K_{M_1...M_{D-5}}$.
This gives rise to the extra
6-brane charge in $D=11$ and the extra 5-brane charge in both type II theories
in $D=10$. The
algebras also suggest that there should be a 9-brane in $D=11$ and a 9-brane in
both the IIA and
IIB theories.
These extra 9-branes can be formally thought of as
coupling to a non-dynamical 10-form potential,  which   could arise   as an
auxiliary field
that is analogous to the 4-form in four-dimensional supergravity which couples
to 3-branes
that was discussed in [\Ovr]. In the IIB theory, the 10-form to which the
Dirichlet 9-brane couples
is associated with the RR sector, suggesting that the new (1,0) 9-branes in the
IIA and IIB theory be
associated with the NS-NS sector. The $D=11$ 9-brane gives rise to the 8-brane
and the
9-brane of the IIA theory, while the NS-NS and the RR 9-branes of  the type IIB
theory   combine
to form
$(p,q)$ 9-branes, with the
$SL(2,\Z)$ U-duality
acting on the $(p,q)$ 9-brane charge lattice.
This will be discussed further in section 6.

 If the $D$-momentum $P_M$ were treated in the same
way as the other charges, it would be split into the spatial momentum $P_i$
which is analogous to a
1-brane charge, and the energy $P_0$, which could be dualised to a $D-1$ brane
charge
$\hat P_{i_1...i_{D-1}}$. The fact that states carrying momentum in
compactified dimensions are
U-dual to states obtained by wrapping branes round the  compactified dimensions
[\HT] means that for
some purposes states carrying the gravitational charge $P_M$ should be treated
on the same footing
as states carrying brane charges. The
BPS states associated with the charge $P_i$ are
pp-waves moving in the $i$'th dimension with  momentum $P_i$ and, as discussed
in the introduction,
these can be regarded as strings or branes for some purposes. Indeed, if the
$i$'th dimension is
compact, the pp-wave with momentum $P_i=n_i $ say is T-dual to a string winding
around the $i$'th
dimension with winding number or string charge $Z_i=n_i$ [\PPdual]. The fact
that the energy appears
to be associated with a $D-1$ brane seems natural from a 12-dimensional
perspective, as will be
discussed in section 7.

\chapter{Magnetic Charges in Gravity}

We have seen that the $IIA,IIB$ and $D=11$ superalgebras have an extra
$(D-5)$-form charge
$K_{i_1...i_{D-5}}$ that it would be natural to associate with a $D-5$ brane.
On dimensionally reducing, the
$D=11$ 6-form charge gives a 6-form charge and a 5-form charge in $D=10$.
The 6-form charge is the magnetic charge for the RR vector field of the type
IIA theory, which
comes from the $g_{\mm 11} $ components of  the $D=11$ metric (where the
reduction is in the
$x^{11}$ direction). Thus the $D=10$ 6-form charge is the magnetic charge of
the   graviphoton, which
arises from Kaluza-Klein monopole solutions of the $D=11$ theory [\PKT], so
that the new
6-form charge in $D=11$
should   be non-trivial in  Kaluza-Klein monopole space-times. The standard
Kaluza-Klein monopole
space-time in
$D$ dimensions  is
$N_4\times \R^{D-5,1}$ where $N_4$ is   4-dimensional self-dual Euclidean
Taub-NUT space (or its
multi-centre generalisation [\Haw]) and
$ \R^{D-5,1}$ is $D-4$ dimensional Minkowski space. This is a solution of
string
theory with $D=10$ and of $M$-theory with $D=11$, since  self-dual
(multi-)Taub-NUT space
is hyperkahler and so is the target space of a (4,4) superconformal sigma-model
for which the
beta-functions vanish [\AGH].
The NUT charge [\RS,\AS,\GHI] of the Taub-NUT space  is related to the magnetic
charge of the
graviphoton on dimensionally reducing [\SGP], so that we learn that the 6-form
charge in $D=11$ is
related to NUT charge. However, the NUT charge is defined for the  4
dimensional space $N_4$, and can
be readily generalised to $D\ge 4$ dimensions, while the $D-5$ form charge $K$
only exists in $D\ge
5$ dimensions, so that it cannot be precisely the $D$-dimensional
generalisation of NUT charge,
although it  is closely related to this.

Consider the Kaluza-Klein reduction of gravity from $D$ to $D-1$ dimensions.
The  $D$-dimensional metric
$g_{MN}$ gives a metric $g_{\mm \nn}$, a vector field $A_\mm$ and a scalar
$\phi$ in $D-1$
dimensions. The $D$-momentum
reduces to the $D-1$ momentum $P_\mm$ and the electric charge $Q$.
In $D-1$ dimensions, the magnetic charge for a Maxwell field $A_\mm$ is a
  $(D-1)-4=D-5$ form, which should have a $D$-dimensional origin. Postulating a
  charge
$ K_{i_1...i_{D-5}}$ in $D$ dimensions,
the reduction to  $D-1$ dimensions then gives a $D-5$ form which can be
associated with the
magnetic charge of the graviphoton $A_\mm$ and a
$D-6$ form $K_{m_1...m_{D-6}}$  (where $m=1,...,D-2$ labels   the spatial
dimensions in $D-1$
dimensions) which corresponds to
the   K-charge
$(D-1)-5$ form in
$D-1$ dimensions.
Thus the gravitational charges $P_1, K_{D-5}$ in $D$ dimensions
give the gravitational charges $P_1, K_{(D-1)-5}$ in $D-1$ dimensions, together
with
the electric and magnetic charges $Q,Z_{(D-1)-4}$ for the graviphoton $A_1$
(where the subscript
denotes the degree of the form).

We now turn to the definition of  the gravitational charges  $P,K$.
We shall be interested in $D$-dimensional space-times $M$  for which a
spacelike slice   is
bounded by a $D-2$ dimensional \lq surface at infinity', $\sss_{D-2}$.
For asymptotically flat spaces, this is a $D-2$ sphere, $\sss_{D-2}=S^{D-2}$,
while for
toroidally wrapped
$p$-brane space-times it is $\sss_{D-2}=T^p\times S^{D-p-2}$.
In particular, for a $D-4$ brane it is $\sss_{D-2}=T^{D-4}\times S^{2}$.
For KK monopole space-times, it is a twisted version of these $D-4$ brane
asymptotics, in which one
of the toroidal directions is a fibre for a circle bundle over $S^2$. For a
single monopole, this
gives the Hopf fibration of $S^3$ as a circle bundle over $S^2$, so that
topologically
$\sss_{D-2}= T^{D-5} \times   S^3 $.  The $S^3$ has a squashed metric -- as a
radial coordinate $r \to
\infty$, the size of the Hopf fibres tends to a constant while the area of the
2-sphere increases
as $r^2$.
For the $s$-monopole generalisation, the surface at infinity is $\sss_{D-2}=
T^{D-5} \times  \ti
\sss_3
$
 where $\ti \sss_3 $  is
the (squashed) 3-sphere for $s=1$ or a (squashed) Lens space $L(s,1)=S^3/\Z_s$
for $s>1$.
(The  Lens space $L(s,1)$ is defined by
regarding $S^3$
as the points $(z,w)$ in $\C^2$ with $\vert z\vert ^2+\vert w\vert ^2=1$ and
identifying $(z,w)$ with
$(e^{2\pi i/s} z, e^{2\pi i/s} w)$.)
 We shall for now
suppose that
$\sss_{D-2} $ is compact, but later we will consider the limit in which the
torus $T^p$ becomes
large, so that $\sss_{D-2}$ tends to $\R^p\times S^{D-p-2}$; in this latter
case it will become
necessary to define the charge densities per unit $p$-volume instead of the
total charges, which
become infinite.

The KK monopole space $N_4\times T^{D-5}\times \R$ can be thought of as a
wrapped $D-5$
brane or as a twisted form of a $D-4$ brane, as
$N_4$ (with a set of fixed points or NUTs removed) has the structure of a
circle
bundle over a base
$\sss$, so that locally it has the wrapped $D-4$ brane structure $\sss\times
T^{D-4}\times \R$.
Indeed, as we shall see, it has a $D-5$ form
charge, which is usually associated with a $D-5$ brane, whereas it has the
number of translational
zero modes appropriate for a $D-4$ brane, as will be seen in section 8.
 On dimensionally reducing with respect to the toroidal directions, the
charges in the resulting $D-p$ dimensional space
correspond to   charge densities per unit $p$-volume.
For the KK monopole, the Hopf fibre  cannot become non-compact without
introducing irremovable
Dirac string singularties in the metric, although the connection and curvature
remain well-defined.
(It was suggested in [\AS] that such singularities might be acceptable for some
purposes, and it
would be interesting to explore their consequences for string theory.)   For KK
monopoles, one can
divide by the length of the Hopf fibre to define densities, as for a $D-4$
brane.

Suppose that there is an asymptotic region in which the  space-time $( M,
g_{MN})$ becomes
close to a background space-time $(\bar M,\bar g_{MN})$, with $g_{MN}=\bar
g_{MN}+h_{MN}$ with
$h_{MN}$ asymptotically small. In the case of $p$-brane space-times, we shall
consider the case in
which
$\bar M$ is $\R^{D-p-1,1}\times T^p $ with a flat metric $\bar g_{MN}$, with
$h_{MN}$ falling off
as  $O(r^{D-3})$ as $r \to \infty $, where $r$ is a radial coordinate on
$\R^{D-p-1}$, so that
for large $r$, $r=constant$ defines a $D-p-2$ sphere of area
$r^{D-p-2}\www_{D-p-2}$, where
$\www_{D-p-2}$ is the area of the unit $D-p-2$ sphere.

For the KK monopole, the question of boundary conditions is more subtle. We can
take
$\bar M=\R^{4,1}\times T^{D-5} $ as Taub-NUT is topologically $\R^4$, while for
the multi-monopole
configurations $\R^4$ is replaced by $\R^4/\Z_s$. (For $s>1$, it is sometimes
convenient to work
with the covering space $\R^4$ instead of $\R^4/\Z_s$, and factor by $\Z_s$ at
the end.)
 There are two approaches to
choosing the background metric $\bar g_{MN}$. One is to fix a   topological
class of metrics, and consider only metrics within that class.
 The natural choice
of background in this case is $\bar N_4\times T^{D-5}\times \R$ where $\bar
N_4$
is   self-dual  Taub-NUT   space or the $s$-centre self-dual generalisation,
and to consider the
class of  spaces asymptotic to this, with the   boundary surface at space-like
infinity given by
$\sss_{D-2}=T^{D-5}\times S^3/\Z_s$. On dimensional reduction, this corresponds
to fixing the
magnetic charge, and considering only spaces with that magnetic charge.
In discussing electromagnetism,
   however,
it is more convenient for some purposes to allow all possible magnetic charges,
even though these
correspond to topologically distinct $U(1)$ bundles, and compare these to the
vacuum which is the
trivial bundle with zero magnetic charge.
Here, this corresponds to allowing   boundary conditions with
$\sss_{D-2}=T^{D-5}\times S^3/\Z_s$
for all values of $s$, and comparing with the flat background metric $\bar
g_{MN}$ of the
Kaluza-Klein vacuum $T^{D-4}\times \R^{3,1}$,
with boundary   at space-like infinity given by
$\sss_{D-2}=T^{D-4}\times S^2$.
In much of what follows, either choice of boundary conditions could be used,
and the formulae give
the difference between the charges $P,K$ of the space-time $(M,g)$ and the
charges $\bar P,\bar K$
of the  background $(\bar M,\bar g)$. However, it will be convenient to present
the results with
respect to the KK vacuum, for which we take $\bar P=0,\bar K=0$.

Introducing frames $e_M^A$ with $e_M^Ae_N^B\eta _{AB}=g_{MN}$ etc and the
corresponding torsion-free
spin-connection $\ww_M{}^{AB}$, Nestor's expression for the ADM momentum $P_M$
[\Nes] can be
generalised to $D$-dimensions to give
$$
P[u]={1\over 2 \www_{D-p-2}}\int _{\sss_{D-2}} u ^{[M} e^N_Be^{P]}_{C} \ddd
\ww_P
{}^{BC} d\sss_{MN}
\eqn\Pu$$
Here $\ddd \ww_P {}^{BC}=\ww_P {}^{BC}- \bar \ww_P {}^{BC}$ is
the difference between the connection on $M$ and that on $\bar M$ and so is an
an asymptotic tensor, so that \Pu\ is covariant.
 The vector $u$ is an asymptotic Killing vector, so that
as $r \to \infty$, $u$ tends to a Killing vector of $\bar M$.
Then $P[u]$ is the
 conserved charge corresponding to $u$. If $\bar M$ has
translational   Killing vectors $ k^a_M$ labelled by some index $a$, then
$u_M=u_a k^a_M$ for some functions $u_a$ tending to constants $\bar u_a$ as $r
\to \infty$. Then
$P[u]$ can be written as $P^a\bar u_a$,
defining the ADM $D$-momentum $P^a=P[k^a]$. In the simplest cases in which
there is asymptotic
flatness, the  index $a$ can be identified with the index $A$ labelling the
asymptotic frames (see
[\Hcmp]). This is equivalent to other forms of
the ADM momentum [\Nes] and is properly the difference between the
ADM momentum and that of the background. It
 can be written in form notation
as
$$
P[u]={1\over 2\www_{D-2}} \int _{\sss_{D-2}} *(e^A_\wee e^B _\wee u)_\wee \ddd
\ww_{AB}
\eqn\abc$$
where $e^A=e^A_Mdx^M$, $\ddd \ww_{AB}=\ddd \ww_{MAB}dX^M $ and $u=u_Mdx^M$.
If $p$ toroidal directions are Killing, so that $\sss_{D-2}=\ti
\sss_{D-p-2}\times T^p$, then
the momentum is proportional to the
volume
$V_p$ of the
$T^p$,
$$P[u]=V_p \tilde P[u],\qquad
\tilde P[u]
={1\over 2 \www_{D-p-2}}\int _{\ti \sss_{D-p-2}} u^{[M} e^N_Be^{P]}_{C} \ddd
\ww_P {}^{BC}
d\sss_{MN}
\eqn\abc$$
   defining the density
$\ti P[u]$ which can be written as an integral
over the surface $\ti \sss_{D-p-2}$ at transverse spatial infinity [\GHT]. Then
the density is well-defined
in the decompactification limit in which $V_p \to \infty$, defining the density
of an infinite brane
space-time. This can be generalised to the case in which the toroidal
directions correspond to
asymptotic Killing vectors, in which case
$V_p$ is the asymptotic volume
of the $p$-torus [\GHT].

Consider a $D-5$ brane space-time, so that the surface at infinity is
$\sss_{D-2}=
\ti \sss _3\times  T^{D-5}$, and let
  $v$ be a   $D-5$-form tending asymptotically to the volume form $\bar v$ on
$T^{D-5}$.
As we shall see, calculating the supersymmetry algebra in such a space gives a
K-charge in general,
and the expression for this charge that arises motivates the following
definition.
 The $D-5$-form charge $K$
$$K[v]={1 \over 5!}
K_{i_1...i_{D-5}}v^{i_1...i_{D-5}}
\eqn\abc$$
is given by
$$K[v]
={1\over 16\pi ^2} \int
_{\sss_{D-2}} \ww _\wedge v
\eqn\abc$$
where $\ww$ is the 3-form
whose components are the totally antisymmetric part of the spin-connection
(minus the background),
$\ddd
\ww_{[ABC]}$,
$$\ww= {1\over2}
e^B _\wee e^A_\wee \ddd \ww_{AB}\eqn\abc$$
Again,
this is the difference between the K-charge and that of the background, and
if the toroidal directions are asymptotically Killing,
$$K[v]= V_p \ti K[v], \qquad \ti K[v]= {1\over 16\pi ^2}\int_{\ti \sss _3} \ww
\eqn\abc$$
If $\ti \sss _3$ is a circle bundle with fibre generated by an asymptotic
Killing vector and if the
length of the fibre at infinity is a constant over $S^2$ and given by $2\pi R$,
then it is
convenient to define the densities
$$\hat P[u]  = {1 \over R} \ti P[u], \qquad \hat K[v]  = {1 \over R} \ti K[v]
\eqn\abc$$

We now show that the charge $K$ indeed occurs in the superalgebra.
We shall suppose that the background space $\bar M$ is supersymmetric, with
Killing spinors
$\aa_0$, satisfying $\ti \nabla \aa_0=0$ (where the supervariation of the
gravitino in the
background $\bar M$ is $\dd \psi _M=\ti \nabla _M \ee$); if
$\bar M$ is flat, a frame can be chosen in which each Killing spinor
$\aa_0$ is constant.
We shall be interested in    asymptotic Killing spinors $\aa$ on
 $M$, i.e.   spinors
on
$M$ tending sufficiently fast to a Killing spinor $\aa_0$ of $\bar M$, $\aa =
\aa_0+O(1/r^p)$ as $r
\to
\infty$ where $p=(D-2)/2$. There is a supercharge $Q[\aa]$ for each commuting
Majorana asymptotic Killing spinor $\aa$ on
 $M$, i.e. for each spinor
on
$M$ tending sufficiently fast to a Killing spinor $\aa_0$ of $\bar M$, $\aa =
\aa_0+O(1/r^p)$ as $r
\to
\infty$ where $p=(D-2)/2$.
  The supercharge is, to lowest order in the gravitino $\psi_P$,
$$Q[\aa]={1\over 2 \www_{D-2}}\int _{\sss_{D-2} }\bar\aa \ggg^{MNP}\psi _P \, d
\sss_{MN}
\eqn\abc$$

For asymptotic Killing spinors to exist, it is necessary that the spin
structure  on $M$ agrees with
that on
$\bar M$ asymptotically.
Consider a background space-time $\bar M$ which is the $D=5$ KK vacuum
$\R^{3,1}\times S^1$. This has two spin
structures, as spinors can be periodic or anti-periodic on $S^1$.
With the periodic spin  structure, the vacuum has Killing spinors, and spaces
$M$ tending to $\bar M$ asymptotically
and whose  spin  structure agrees with that of $\bar M$ asymptotically will
have asymptotic Killing spinors. Then there is a
positive mass theorem and a Bogomolnyi bound for such spaces, and the vacuum is
stable against fluctuations which have the
same spin structure asymptotically as the vacuum.
On the other hand, with the anti-periodic spin  structure, the vacuum has no
Killing spinors and is not supersymetric,
and spaces $M$ tending to $\bar M$ with anti-periodic spinors at infinity will
have no asymptotic Killing spinors
and so need not have positive mass. Indeed, there are negative mass
configurations with these asymptotics, leading to the
instability of the KK vacuum [\Winstanton]. The 5-dimensional Euclidean
Schwarzschild space tends asymptotically
to the Euclidean vacuum $\R^4\times S^1$ with anti-periodic spin-structure, and
can be continued to a Lorentzian
expanding bubble solution responsible for the decay of the KK vacuum
[\Winstanton]. Thus stability, positivity of the mass
and a BPS bound can only be established for spaces with supersymmetric boundary
conditions.  With   non-supersymmetric
asymptotics, the loss of supersymmetry means that there is nothing to prevent
the ADM  mass from becoming negative, and  the
 vacuum is   unstable.  For example, the space
$Y\times \R$, where $Y$ is $D=4$ Euclidean Schwarzschild space,   does not
admit asymptotic Killing spinors, since the unique
spin structure on
$Y\times \R$ requires the spinors to be anti-periodic in the $S^1$ at infinity
generated by
translations in the periodic Euclidean Schwarzschild time.
On the other hand, if $Y$ is Euclidean Taub-NUT (with mass $M$ and NUT
parameter $N$), the spinor structure is periodic and
there are asymptotic Killing spinors. If $M=\vert N\vert$, $Y$ is self-dual and
has covariantly constant spinors, while
if $N=0$, $Y$ reduces to Euclidean Schwarzschild space and
the
limit $N\to 0$ leads to a change in topology and spin structure.
In what follows, we shall consider only spaces with asymptotic Killing spinors,
and will
prove stability and a Bogomolnyi bound within this class of spaces.

For commuting asymptotic Killing spinors $\aa,\bb$, the supercharges satisfy
the algebra [\Hcmp]
$$ \{ Q[\aa], Q[\bb] \}
= {1\over 2\www_{D-2}}\int _{\sss_{D-2}} E^{MN}(\aa,\bb)d\sss_{MN}
\eqn\salg$$
in a bosonic background (otherwise there are corrections of second order in
fermion fields) where
$$E^{MN}(\aa,\bb)={1\over 2}\bar \aa \ggg^{MNP}\hat \na _P\bb + c.c.
\eqn\nesis$$
where $\hat \na$ is the supercovariant derivative occurring in the gravitino
supersymmetry
transformation $\dd \psi_M=\hat \na _M\ee$ and $c.c.$ denotes complex
conjugate. In purely
gravitational backgrounds,
$\hat
\na$ is the usual gravitational  covariant derivative $\na$.  Using the
asymptotic behaviour of the
spinors,
\salg\ can be rewritten  in terms of the charges
$P,K$ as
$$ \{ Q[\aa], Q[\bb] \}
= P[\bar \aa \ggg ^M \bb] +K[*(\bar \aa \ggg ^{M_1 ... M_{D-5}}\bb)] +\sum _p
{1\over p!}Z_{M_1\dots M_p}
\bar \aa \ggg ^{M_1 ... M_{p}}\bb
\eqn\abc$$
where the $Z_{M_1\dots M_p}$ are the electric and magnetic
 anti-symmetric tensor charges, discussed in section 2; the values of $p$ that
occur depend on the
algebra, and are given in section 2 for $D=10,11$. If $\aa=\bb$, \nesis\
becomes the Nestor tensor
[\Nes] which can be used to derive a BPS bound for the mass in terms of   $K$
and
the anti-symmetric tensor charges, as in [\GH,\GHT].\foot{In certain cases,
such as the
IIB algebra \tob\ and the $N=2$ algebra in $D=4$, but not the $D=11$ algebra or
the $IIA$ algebra,
there is also a 3-form charge arising in the gravitational sector; this will be
discussed further
elsewhere.}  This then implies similar bounds for the $p$-brane densities $\ti
P,\ti K$
and for the KK monopole densities $\hat P,\hat K$.
The bound requires that the matter system satisfies a positive energy
condition, and that
the energy density is bounded below by the appropriate charge densities, as in
[\GH,\GHT,\Witpos]. It also
requires the existence of
asymptotic Killing spinors satisfying the \lq Witten  equation'
[\GH,\GHT,\Witpos]; when
asymptotic Killing spinors exist, they can usually be chosen to satisfy the
Witten  equation
and this has been proven in four dimensions for asymptotically flat spaces
[\Wex].
 The requirement of the existence of
asymptotic Killing spinors implies that the bound only applies to systems
satisfying supersymmetric
boundary conditions, which in particular requires that the spin structure agree
asymptotically with
that of a supersymmetric space.

\chapter {Bogomolnyi Bounds in Five Dimensions and Four Dimensions}

In the case $D=5$, $K$ is a scalar and so occurs as a central charge in the
$D=5$ superalgebra.
The $N$ extended superalgebra in $D=5$ is
$$\eqalign{
\{Q_\alpha^a,Q_\beta^b\} &=\, \www^{ab}\big( \Gamma^MC\big)_{\alpha\beta} P_M +
\www^{ab} C _{\alpha\beta}K
\cr
& +
\big( \Gamma^MC\big)_{\alpha\beta} Z_M^{ab} +
\www^{ab} C _{\alpha\beta}Z^{ab}  + \big(
\Gamma^{MN}C\big)_{\alpha\beta}Z_{MN}^{ab}
\cr}
\eqn\fialg$$
Here $N$ is even, the supercharges $Q_\alpha^a$ with $\aa=1,\dots ,4$ and
$a=1,\dots N/2$ are
symplectic Majorana spinors and $\www^{ab}$ is the symplectic invariant of
$USp(N)$. The central
charges
$Z^{ab}$ satisfy $Z^{ab}=-Z^{ba}$ and
$\www_{ab}Z^{ab}=0$, the charges $Z_M^{ab}$ satisfy $Z_M^{ab}=-Z_M^{ba}$ and
$\www_{ab}Z_M^{ab}=0$
while $Z_{MN}^{ab}=Z_{MN}^{ba}$. The central charges $Z^{ab}$  arise as 0-brane
charges, the $Z_i^{ab}$
  arise as string charges, and there are  also 2,3 and 4-brane charges given by
$Z_{ij}^{ab}$, $\hat Z_{ijk}^{ab}$ and $\hat Z_{ijkl}^{ab}$, respectively.
 The gravitational charges are the 5-momentum $P_M$ and the central charge
$K$. For example, for $N=8$ in $D=5$, the supermultiplet contains 27 abelian
gauge fields giving 27 electric charges
$Z^{ab}$ and 27 magnetic string charges $Z_i^{ab}$.
On reduction to $D=4$, the 28 electric central charges come from $P_5$ and
$Z^{ab}$, while the 28 magnetic central charges come from
$Z_5^{ab}$ and $K$. The $D=5,N=8$ algebra with central charge $K$ was
considered in this context in
[\GP]. The
 application of the positive mass theorem to
$D=5$ KK monopoles was also considered in [\LS-\BM].

The algebra \fialg\ gives   BPS bounds for $p$-branes, in
the usual way [\GH,\DGHR,\GHT,\Witpos,\GP] for spaces with supersymmetric
boundary conditions with matter satisfying a
\lq local BPS condition' that its  energy density is bounded below by an
expression depending on charge densities.
 For 0-branes, or for asymptotically flat or KK (multi-) monopole space-times,
in
$D=5$, the mass satisfies
$$M \ge   {\vert \ll_n\vert}
\ek
where the $\ll_n$, $n=1,...,N/2$ are the skew eigen-values of
$Z^{ab}+K\www^{ab}$.
In particular, this implies
$$M\ge \vert K\vert
\ek
with equality for   space-times $\R\times N_4$ where $N_4$ is self-dual or
anti-self-dual.
Dividing by the size of the $S^1$ fibre, this gives
$$ \hat M \ge \vert \hat K\vert
\ek
and as we shall see in   section 5, $\hat K$ can be identified with the NUT
charge $N$.
Dimensionally reducing with respect to the time coordinate to $N_4$ gives the
bound
$$\hat M \ge \vert N\vert
\ek
for four-dimensional   Euclidean geometries with asymptotic Killing spinors
with equality for spaces admitting covariantly constant spinors, i.e. for
self-dual geometries.
This can be obtained directly by considering
the dimensional reduction of \salg\ to $N_4$ (i.e. reducing in the time
direction) given by
$$  {1\over 2\www_{3}}\int _{\ti \sss_3} E^{MN}(\aa,\bb)d\sss_{MN}
\eqn\salgo$$
where
$$E^{MN}(\aa,\bb)={1\over 2}\bar \aa \ggg^5\ggg^{ NP}\hat \na _P\bb + c.c.
 \eqn\nesios$$
The bound then follows from standard arguments using $E^{MN}(\aa,\aa)$. For ALF
boundary conditions,
${\ti \sss_3}$ is a circle bundle over $S^2$.
The quantity $\hat M$ is the ADM momentum $P[k]$ corresponding to the
asymptotic Killing vector
$k$ generating the
$S^1$ fibre.
Including electric and magnetic charges $Q,P$, this becomes
$$ \hat M^2 + Q^2 \ge N^2 + P^2
\ek
This can formally be continued to Lorentzian signature metrics. Wick rotating
the $S^1 $ fibres
(i.e. in adapted coordinates in which $k=\pa/\pa y$, we Wick rotate the
coordinate $y \to i t$)
an argument similar to that of [\GH] gives
$$ \hat M^2 + N^2 \ge Q^2 + P^2
\ek
with equality only for spaces admitting supercovariantly constant spinors
$\aa$, $\hat \na \aa=0$.  Such a bound was suggested in [\Ort].

\chapter {Evaluation of $K[v]$ and Relation to NUT Charge.}

Consider a     $D$ dimensional manifold $M$ with an isometry generated by a
space-like Killing vector
$k= {\pa /\pa y}$.
The main examples considered here will be $M=N_4\times \R^{D-5,1}$ where $N_4$
is   4-dimensional
 Taub-NUT space or its multi-centre generalisation.
The isometry defines a fibering
$\pi:M \setminus {\cal F} \rightarrow B$
where ${\cal F}$ is the fixed-point set of the action of $G$,
so that $B$ is the space of non-degenerate orbits.
The metric $G_{MN}$ on $M$ induces a Lorentzian signature metric
$$ g_{MN}=G_{MN}-V^{-1}k_Mk_N
\eqn\abc$$
on $B$, where $V= k^Mk_M$.
The metric $G_{MN}$ on $M$ can be written as
$$ds^2=V(d y + A_\mu d x^ \mu)^2 +V^{-1}g_{\mu \nu}d x^\mm
d x^\nn
\eqn\remet$$
where $V,A_\mu$ and $g_{\mu \nu}$ do not depend on the coordinate $y$.
The vector field   $A_\mm$
  is defined up to a gauge transformation
$$A_\mm\rightarrow A_\mm+  \partial_ \mm \rho
\eqn\varo$$
as such a change  can be absorbed into the coordinate transformation
$$ y \rightarrow y - \rho
\eqn\abc$$
The invariant field-strength
$$F_{\mm\nn}= \partial_{[\mm} A_{\nn]}
\eqn\twi$$
is the {  twist} or {  vorticity} of the vector field $k$ and can be written
covariantly in $D$
dimensions as
$$F_{MN}= V^{-1}g  _M^{\ P}g_N^{\ Q}\nabla_{P} k_Q
\eqn\abc$$

On dimensional reduction with respect to the Killing vector $k$, the
$D$-dimensional metric
$g_{MN}$ gives a metric $g_{\mm\nn}$, a graviphoton $A_\mm$ and a graviscalar
$V$ on the $D-1$
dimensional space $B$. The kinetic term for the metric $g_{\mm \nn}$ is not
conventionally
normalised; the Einstein metric is given by $g^E_{\mm \nn}=V^\aa g_{\mm \nn}$
where $\aa=-1/D-2$.
Note that the two metrics $g_{\mm \nn}$ and $g^E_{\mm \nn}$ lead to different
ADM momenta, $P_\mm$
and $P^E_\mm$ respectively [\Bom].
 The vector field $A_\mm$ is an abelian gauge field on
$B$, which will have electric charge $q$ and a magnetic charge given by a $D-5$
form $Z_{m _1...
m_{D-5}}$, where $x^m$ are the spatial coordinates on $B$, with $m=1,\dots,
D-2$.
If the boundary of a spatial slice of $B$
is $\R^{D-5}\times \sss^2$ or  $T^{D-5}\times \sss^2$
for some compact 2-space $\sss^2$,
then
$$Z.v ={1\over 4\pi}\int_{\sss^2} F
\eqn\abc$$
 where $v$ is the volume-form on $\R^{D-5}$.
If $M=N_4\times \R^{D-5,1}$, then
the NUT charge defined in [\RS,\AS,\GHI]  can be generalised to a $D-5$ form
charge $N_{\mm_1
...\mm_{D_5}}[k]$ for $M=N_4\times \R^{D-5,1}$ (or for the Euclidean space
$M=N_4\times \R^{D-5}$)
given by
$$
N[k].v = {1\over 8\pi}\int_{\sss^2} F= {1 \over 2} Z.v
\eqn\nis$$
(The factor of $8\pi$ in \nis\ is included to agree with the normalisation in
[\RS,\AS,\GHI].) If $y$
is  periodic with period
$P$,
$y
\sim y+P$, then the length of the orbit of
$k$ at
$x^\mm$ is $\sqrt {V(x)} P$ which tends to $2\pi R$, say, on $\sss$. The
electric charge $q$ of the
graviphoton is proportional to the momentum in the $y$ direction
$ q \propto{1\over  R} P^E[k]
$,
while, as we shall see, the magnetic charge is
$ Z.n\propto {1\over 2R}K[n]
$.

The momentum $P[k]$  and the corresponding
 NUT charge satisfy a Dirac quantization condition   and the relation of
the NUT charge  to mass is similar to the relation of
 magnetic charge   to
electric charge. Two     generalisations of the NUT charge or dual mass to
spaces without isometries
are given in [\RS,\AS]. However, both invoke special structures at   infinity,
whereas the definition of
the charge $K[v]$ is general and  needs no such structures.

Let $e_M^A$ and $\hat e _\mm^a$ be frames on $M,B$ respectively, satisfying
$$
e_M^Ae_N^B\eta_{AB}=G_{MN},\qquad \hat e_\mm^a \hat e_\nn^b
\eta_{ab}=V^{-1}g_{\mm\nn}
\eqn\abc$$
A suitable choice of one-forms $e^A=\{e^y,e^a\}$ is
$$
e^y=V^{1/2}(dy+A_\mm dx^\mm),\qquad
e^a=\hat e^a_\mm dx^\mm
\eqn\abc$$
 The corresponding spin-connection one-forms $\ww ^A{}_B,\hat
\ww^a{}_b$ satisfy
$$d e^A+\ww ^A{}_B{}_\wee e^B=0,
\qquad d\hat e^a+\hat \ww^a{}_b{}_\wee \hat e^b=0
\eqn\abc$$
so that
$$\eqalign{\ww^y{}_a&= \2 V^{-1} V_a e^y + V^{1/2} F_{ab} e^b,
\cr
\ww^a{}_b&= \hat \ww^a{}_b-  V^{1/2} F_{ab} e^y
\cr}
\eqn\abc$$
where $V_a=\partial _\mm V e^\mm_a$ and $F_{ab}=e^\mm_ae^\nn_bF_{\mm\nn}$ with
$F_{\mm\nn}=2\partial
_{[\mm}A_{\nn]}$.
Then the 3-form corresponding to the totally antisymmetric part of the
spin-connection is
$$\ww= F_\wedge k +\hat \ww
\eqn\wis$$
where
$$k=V (dy+A_\mm dx^\mm)=k_M dx^M, \qquad \hat \ww = e^b_\wedge e^a \hat \ww
_{ab}
\eqn\abc$$
and $F=\2 F_{\mm\nn} dx^\mm_\wedge dx^\nn
$. In what follows, we shall consider only the case in which the contribution
$\hat \ww$ to \wis\
from the geometry of
$B$ is zero. Such contributions could have arisen, for example, if there were
several Killing vectors
giving rise to several magnetic charges, and their treatment is a
straightforward generalisation
of the case considered here. (Alternatively, instead of considering asymptotics
with respect to a
flat background, we could consider  asymptotics with respect to a background
with $\bar \ww ^a{}_b=
\hat \ww ^a{}_b$, so that $\ww\equiv e^b_\wedge e^a _\wedge(  \ww _{ab}-\bar
\ww _{ab})=
F_\wedge k$.)

The isometry defines a fibering
$\pi:\ti \sss_3 \setminus {\cal F}' \rightarrow \sss ^2$
where $\sss_{D-2}=T^{D-5}\times \ti \sss_3$ or $\sss_{D-2}=\R^{D-5}\times \ti
\sss_3$
is the
boundary of a spatial slice of $M$ and $T^{D-5}\times\sss ^2$ or
$\R^{D-5}\times\sss ^2$  is the boundary of a spatial slice of $B$,
and ${\cal F}'$ is the fixed-point set of the isometry action. Then $F_\wee k$
is proportional to
the  volume form of $\ti \sss_3$, so that $\int \ww$ is a winding number.
{}From \wis\ and the fact that $k$ is a Killing vector, it follows that
$$\ti K={1\over 16 \pi ^2}
\int _{ \ti \sss_3} \ww={1\over 16 \pi ^2}
\int_{S^1} k \int _{ \ti \sss } F
= {R\over 8\pi} \int _\sss F = RN
\eqn\abc$$
where $2\pi R=\int k$ is the length of the fibre at infinity.
In particular, this implies that
$$\hat K=N
\ek
 In  the case of Taub-NUT, $ \ti \sss_3$ is
$S^3$ and $\sss^2=S^2$, with $\pi$ the Hopf fibration. This case will now be
examined in
detail.

The Euclidean Taub-NUT metric
is  of the form \remet; taking   $x^\mm$ as spherical polar coordinates $r,
\th, \phi$, it can be
written as
$$ds^2=V(d y+ A_\phi  d \phi )^2 + V^{-1}dr^2 +r^2 f(d \theta^2 + \sin ^2
\theta
d \phi^2)
\eqn\nut$$
where
$$ V=1-{2(Mr+N^2) \over r^2+N^2},\qquad A_\phi=
4N \sin^2  {\theta \over 2},
\qquad f= 1-{2M \over r}+{N^2 \over r^2}
\eqn\tubis$$
The Killing vector $k=\pa/\pa y$ can be used to define a component of momentum
$P[k]$ and a
 NUT charge $N[k]$  given by the parameters $M,N$ respectively.
The limit $N \to 0$ gives Euclidean Schwarzschild with \lq Euclidean mass' $M$
(i.e. the mass
resulting from regarding $y$ as the Euclidean time).
 The parameter $M$ is the $k$-component of momentum $P[k]$ and $N$ is the
corresponding NUT charge. The curvature is self-dual if $M=N$ and
anti-self-dual if $M=-N$.
Regarded as a metric on $\R^4$ with spherical polar coordinates $ \theta,
\phi$, this metric has a
Dirac string or wire singularity along the half-axis
$ \theta =\pi$.
This singularity can be removed by introducing a new coordinate
$$y'= y+4N \phi
\eqn\newt$$
The metric becomes
$$ds^2=V(d y'+ A_\phi '  d \phi )^2 +  V^{-1}dr^2 +r^2 f(d \theta^2 + \sin ^2
\theta d \phi^2)
\eqn\nuts$$
with
$$ A_\phi'= -4N \cos^2  {\theta \over 2}
\eqn\omp$$
The field strength $F=dA=dA'$ is well-defined and given by
$$F = 2N\sin \theta d \th _\wee d \phi
\eqn\abc$$
The metric \nuts\ is regular  at $ \theta=\pi$ but not at $ \theta=0$. A
non-singular  metric is obtained by
using $t,r,\theta,\phi$ in the patch
$0 \leq \theta < \pi $ and $t',r,\theta,\phi$ in the patch
$0 < \theta \leq \pi $. In the overlap, the transition \newt\ is   consistent
with the periodicity
of
$\phi$
 if we make the $y$ coordinate periodic with period $8 \pi \vert N\vert $. This
 changes the
topology of the surfaces $r= {\rm constant} $ from $S^2 \times \R$ to $S^3$ and
leads to the
requirement that the energy $E$ of any particle moving in the space-time has to
satisfy the Dirac
quantization condition
$4NE={  integer}$. Then $\psi, \phi,\theta$ are Euler angles on $S^3$ with
$\psi=y/2\vert N\vert $.
The
$S^3$ has a squashed metric: as $r\to
\infty$, the length of the $S^1$ Hopf fibres tends to a constant, $8\pi \vert
N\vert $, while the
area of the
$S^2$ base increases as $r^2$.

The three-form $\ww$ is given by
$$\ww= 2NV \sin \th d \th _\wee d \phi_\wee  dy=2NV \sin \th d \th _\wee  d
\phi _\wee  dy'=4N\vert
N\vert  V\sin \th d \th _\wee d \phi _\wee d\psi
\eqn\abc$$
and is proportional to the volume form  on $S^3$ (and is  well-defined).
Integrating on an
$S^3$ of constant
$r$ and taking the limit
$r\to \infty$ gives
$$
\ti K= {1\over 16 \pi ^2}\int \ww= 4 N\vert N\vert
\eqn\abc$$
The $S^3$ is a circle bundle over the 2-sphere  $\sss$ parameterised by
$\th,\phi$ with fibre
coordinate
$\psi$. The NUT charge \nis\ is
$$ {1\over 8\pi}\int _\sss F={1\over 8\pi}\int _\sss 2N \sin \th d \th d \phi =
  N
\eqn\abc$$
so that they are indeed related by $K=RN$, $\hat
K=N$, where $R$ is the radius of the $S^1$, $R=4\vert N\vert $.

The
self-dual multi-Taub-NUT solution [\Haw] is also of the form \remet, but with
$$ V^{-1}=1+\sum _{i=1}^s {2N_i \over \vert x^\mm-x_i^\mm\vert },
\qquad
F_{\mm\nn}=\ee
_{\mm\nn\rr}\nabla^\rr V^{-1}
\eqn\vis$$
and with
$$g_{\mm\nn}=\dd_{\mm\nn}\eqn\gist$$
and this flat metric is used to raise and lower
three-dimensional indices $\mm,\nn,\dots$.
There are $s$ nuts with parameters $N_i$ at the positions $x_i^\mm$.
The potential $A$ will again have Dirac string singularities in general, but if
all the parameters
$n_i$ are equal to a single value $N$,
then all singularities are removed by identifying $y$ with period $8\pi \vert
N\vert $.
The surface at infinity is the Lens space $L(s,1)$.
As $\vert x^\mm\vert \to \infty$,
$$V \to 1, \qquad F\to 2Ns  \sin \th d \th _\wee d \phi,\qquad \ww \to 2Ns
\sin \th d \th _\wee d
\phi _\wee dy
\eqn\abc$$
so that integrating $\ww$ over the Lens space $L(s,1)$ at infinity gives
$$
\ti K={1\over 16 \pi ^2}\int \ww = {1\over s}{1\over 16 \pi ^2}
\int 2Ns \sin \th d \th d \phi dy
=4 N\vert N\vert
\eqn\abc$$
while the NUT charge is
$$N={1\over 8 \pi  }\int _\sss F={1\over 8 \pi  }\int _\sss 2Ns \sin \th d \th
d \phi =  sN
\eqn\abc$$
so that they are related by $K=RN$, where $R$ is the \lq size' of the fibre
$S^1 /\Z_s$, $R=4
\vert N\vert /s$.

\chapter{Nine-Branes in IIB Theory}

We now turn to the new $9$-branes
that have been proposed for the type IIB theory.
That such branes should be present in the theory follows from duality and
supersymmetry.
It is known that there are D-9-branes with RR charge in the type IIB string,
and that they break half the supersymmetry, so that there should be a
corresponding 9-brane charge
in the IIB superalgebra.
Consider now the action of an $SL(2,\Z)$ duality transformation  on the RR
9-brane.
The 9-brane will either be invariant, or will be mapped into a new 9-brane.
However, we have seen from the superalgebra  that the 9-brane charges that
occur on the right hand side of \tob\ fit into
doublets of   $SO(2)$,
as do the 1-brane and 5-brane charges. For the 1-brane and 5-brane, the $SO(2)$
doublet of charges is related (by a
scalar-dependent transformation) to an $SL(2)$ doublet of charges, and in the
same way the
$SO(2)$ doublet of 9-brane charges $(Z^1,Z^2)$  can be converted to an $SL(2)$
doublet of 9-brane charges $(\bar
Z^1,\bar Z^2)$. The D-9-brane charge $\bar Z^2$ is then not a singlet of
$SL(2,\Z)$, so the D-9-brane is mapped to a
new 9-brane, and the action of $SL(2,\Z)$ on the doublet of 9-brane charges
will give
$(p,q)$ 9-branes with $p,q$ co-prime integers.

 In the type IIB theory, the RR 9-brane occurs as a D-brane and gives rise
to Chan-Paton factors for the fundamental string at weak coupling.
A fundamental string can end on a D-brane, and
the standard configuration with $n$ 9-D-branes
would be with the $n$ branes coincident and each filling $D=10$ space-time.
Thus strings ending on
a 9-brane would be open strings, with each end labelled by  which of the $n$
9-D-branes it ended on,
giving rise to an $n$-dimensional Chan-Paton factor.
  In $D=10$ Minkowski space, charge conservation forbids the presence of any
9-D-branes. However, if one mods out by world-sheet parity reversal $\www$ to
construct an
orientifold, the whole $D=10$ Minkowski space
is a source of 9-brane charge
which can be cancelled by adding precisely 32 9-D-branes, giving rise to the
type I string with
$SO(32)$ Chan-Paton factors.

As the presence of 9-branes is usually not allowed without the presence of some
other
sources of charge, such as an orientifold, any discussion of 9-branes in
general is necessarily rather formal.
 Before discussing the possibility of   generalising  the usual
orientifolds to give sources of $(p,q)$ 9-brane charge so as to give consistent
backgrounds with
$(p,q)$ 9-branes, it will be useful to consider 9-branes and their relations
to other branes further.

The type IIB theory has an $SL(2,\Z)$ symmetry [\HT,\Oopen] which acts on the
complex scalar $\lambda
=
\chi +ie^{-\Phi}$, where
$\Phi$ is the dilaton and $\chi$ is the RR scalar, and the two anti-symmetric
tensor field strengths
$H_{ MNP }^u$ with
$u=1,2$ where
$H ^1=dB^1$ is the NS-NS field strength and $H ^2=dB^2$ is the RR field
strength. The transformations are
$$\ll \to {p \ll +q \over r \ll +s}, \qquad
\pmatrix {H^1 \cr
H^2 } \to \lll
\pmatrix {H^1 \cr
H^2 }
\eqn\slt$$
where $\lll$ is the $SL(2,\Z)$ matrix
$$
\lll=\pmatrix {p &r \cr
q & s}, \qquad ps-qr=1, \qquad p,q,r,s \in  \Z
\eqn\lis$$
The branes of the type IIB theory
  fit into representations of the $SL(2,\Z)$ symmetry. Some are singlets, while
some of the
branes transform as   doublets of the $SL(2,\R)$ symmetry of the classical
supergravity theory
  leading in the quantum theory to a spectrum of branes with charge $(p,q)$
with $p,q$ co-prime
integers. The 3-brane is a singlet and the strings and 5-branes are doublets.
The 7-brane is more subtle, as it couples to the scalar fields which transform
non-linearly under $SL(2,\Z)$. The
solutions of [\GGP] in which the axion ansatz
involves the modular invariant $j$-function are $SL(2,\Z)$ invariant and
lead to singlet 7-brane charges.
Acting with $SL(2,\Z)$ on a 7-brane leaves its charge $Z_{i_1...i_7}$
invariant, but changes the
$SL(2,\Z)$ monodromy
and the couplings to
strings and 5-branes [\Doug].
This is consistent with the superalgebra, which we have seen has   singlet
3-brane and  7-brane charges, and
1-brane and 5-brane charges that are doublets.
The 9-brane charges are doublets, so that there should be  9-branes with
charges $(p,q)$ with $p,q$
co-prime.

Strings carry two one-form charges $\bar Z_i^u $ and 5-branes carry two 5-form
charges
$\bar Z_{ijklm}^u$, which transform as doublets under $SL(2)$;
$$\bar Z_i=\pmatrix {\bar Z_i^1 \cr
\bar Z_i^2 } ,\qquad
\bar Z_{ijklm}=\pmatrix {\bar Z_{ijklm}^1 \cr
\bar Z_{ijklm}^2 } \eqn\sltas$$
  The scalars  can be used to construct a $2\times 2$ matrix ${\cal V}\in
SL(2,\R)$
transforming under
rigid $SL(2,\R)$ from the right and under local $SO(2)$ from the left, so that
under
$SL(2,\Z)$,  ${\cal V}\to {\cal V} \lll ^{-1}$.
Fixing an $SO(2)$ gauge gives an expression for
If ${\cal V}\to {\cal V}_0$ at infinity, for ${\cal V}$ in terms of the complex
scalar $\ll$.
some constant matrix
${\cal V}_0$,
then
$ Z_i ={\cal V}_0\bar Z_i $ and $
Z_{ijklm} ={\cal V}_0\bar Z_{ijklm}$ are $SL(2)$-invariant charges which
however transform as
doublets under the
$SO(2)$ which also acts on the fermions and supercharges; it is these charges $
Z_i,
Z_{ijklm}$ that enter into the
superalgebra \tob.

A similar structure  should hold for the 9-branes. The 2-vector of 9-brane
charges $Z_0 $ (dual to $\hat Z_{i_1...i_9}^{ab}$)
 in the algebra is an $SO(2)$
doublet  and should arise from an $SL(2)$ doublet $\bar Z_0$, with $ Z_0 ={\cal
V}_0\bar Z_0 $. The D-brane with $p=9$ can be
thought of as coupling to a 10-form potential $A^2$ which is an auxiliary field
of the theory, occurring in the RR sector.
(Strictly speaking, the RR construction gives only physical fields, but the
9-brane coupled to the 10-form is related by
T-duality to $p$-branes coupling to RR ($p+1$) form gauge fields.) As the
charges $\bar Z_0$ are an $SL(2)$ doublet, there
should be an $SL(2)$ doublet of 10-form potentials, $A^1,A^2$. The  field $A^1$
is again non-physical, and it seems natural to
attribute it to some generalisation of the usual NS-NS sector to include
auxiliary fields.

At weak coupling, $g \equiv <e^\Phi > \approx 0$, the perturbative states are
described by the  NS-NS or
(1,0) string   while all the other branes are non-perturbative [\HStr]. The
perturbative theory is formulated as the
usual type IIB superstring theory, with a topological expansion in terms of the
genus of the world-sheet of the
NS-NS string. At strong coupling, however, it is the RR or (0,1) string that
gives the states that are  perturbative
in an expansion in $\ti g = 1/g$. The conjectured self-duality of the theory
implies that the strong-coupling
theory is again a type IIB string theory, but  now the formulation should be in
terms of the world-sheet of the
(0,1) string.
The perturbation theory in $\ti g$ is a sum over $(0,1)$ string world-sheets
with the power
of $\ti g$ corresponding to the genus of the world-sheet.
 At weak coupling, the (1,0) string can end on the D-branes carrying RR charge,
which are the 3-brane,
the 7-brane, the (0,1) string, the (0,1) 5-brane and the RR 9-brane, which we
will refer to as
the (0,1) 9-brane.\foot{The (1,0) string can also end on $(n,\pm 1)$ strings,
5-branes and
9-branes, as follows from $SL(2,\Z)$ duality. However, it will be sufficient
here to focus on the
(0,1) Dirichlet branes.}
 In the  strongly coupled theory formulated in terms of the world-sheet of
the (0,1) string, the (0,1) string and (0,1) 5-brane carry charge that appears
in the NS-NS sector
of the dual (0,1) string (as this couples to
$B^2$) while the new D-branes with density proportional to $1/\ti g$ on which
the (0,1) string can end are the
 3-brane,
the 7-brane, the (1,0) string and  the (1,0) 5-brane.
These all carry charge which occurs in the RR sector of the (0,1) string. This
structure is obtained by
acting with the $SL(2,\Z)$ transformation generated by
$$
S= \pmatrix {0 &1 \cr
-1 & 0}
\eqn\soi$$
which interchanges strong and weak coupling.
For example, this takes a (1,0) string ending on $(0,1)$ strings or 5-branes to
a (0,1) string ending on $(1,0)$ strings
or 5-branes.

This should extend to 9-branes as follows. There are two basic 9-branes, with
charges (1,0)  and (0,1).
At weak coupling, the (0,1)
 9-brane   is a D-brane for the weakly coupled (1,0) string
and
carries RR charge.
The strongly coupled theory is formulated in terms of the world-sheet of the
(0,1) string and has
the (1,0) 9-brane as a D-brane  occurring   in the RR sector of the (0,1)
string.
A general  9-brane has two charges $(p,q)$ corresponding to the two 10-forms
$A^1,A^2$ and can be thought of as a bound state
of (1,0) and (0,1) branes. The  two auxiliary 10-form fields fit into an
$SL(2,\Z)$ doublet and
one  occurs in the RR sector of the (1,0) string, while the other occurs in the
RR sector
of the (0,1) string.
(A naive extrapolation of the formulae of [\HStr] suggests that for the
weakly coupled string, the coupling constant dependence of the density of the
NS-NS 9-brane is
$g^{-4}$. This is to be compared with that of the RR 9-brane and other D-branes
which have
densities of order   $g^{-1} $ and  that of the solitonic
NS-NS 5-brane which has density of order $g^{-2}$.
Then on going to strong coupling and using the dual string metric, the (1,0)
9-brane would have density $\ti g
^{-1}$ and the (0,1) 9-brane would have density
$\ti g ^{-4}$.
 However, this is rather formal as there are various problems in
discussing the densities of $p$-branes with $p\ge 7$ in ten dimensions.)

Instead of acting with the $SL(2,\Z)$ transformation \soi\ that takes us from
weak  to  strong coupling, we could act
with a general $SL(2,\Z)$ transformation generated by
\lis. The $(1,0)$ {}$1,5,9$-branes are mapped to $(p,q)$ {}$1,5,9$-branes and
the $(0,1)$ {}$1,5,9$-branes  are mapped to
$(r,s)$ branes, while the transformation \slt\ of $\ll$     leads to a new
coupling constant $\ti g$.
Perturbation theory in $\ti g$ again gives a type IIB string theory, but where
now the $(p,q)$
string is fundamental and   formulated in terms of the $(p,q)$ string
world-sheet.
The D-branes of this dual world-sheet description are the $(r,s)$ 1-branes,
5-branes and 9-branes, together with the
3-brane and the 7-brane.
These couple to potentials that arise in the RR sector of the new fundamental
string.
The $(p,q)$ 5-brane is now solitonic, and other strings and 5-branes occur as
various  bound states.

We turn now to the orientifold construction.
At weak coupling, the type IIB theory is formulated in terms of a fundamental
$(1,0)$ string and has a
symmetry $\www$ which reverses the parity of the $(1,0)$ world-sheet. As the
dilaton is invariant under
$\www$, this is a perturbative symmetry.
The massless bosonic fields of the IIB theory are $g_{MN}, B^1_{MN}, \Phi$ in
the NS-NS sector and
$D_{MNPQ},  B^2_{MN}, \chi$ in the RR sector.
Of these, the ones that are invariant under $\www$ are $g_{MN}, B^2_{MN},
\Phi$. The orientifold of the IIB theory
constructed using $\www$ gives the type I theory [\Sag]. The invariant sector
is formulated in terms of closed type I
strings, with massless bosonic fields
 $g_{MN}, B^2_{MN}, \Phi$. In addition there is an open string sector, which
can in some ways be thought of as a
twisted sector, with $SO(32)$  Chan-Paton factors  arising from 32 RR 9-branes
[\PW].
Acting with $\www$ projects out all branes except the RR string, 5-brane and
9-brane, although there is a sense in
which the closed and open fundamental type I strings can be associated with the
fundamental NS-NS string of the type IIB
theory.

Consider now the strongly coupled type IIB string, formulated as a world-sheet
theory of the (0,1) string.
This should be isomorphic to the usual type IIB string,
and in particular has a symmetry $\ti \www$  which reverses the parity of the
(0,1) string world-sheet.
As $\ti \www$ leaves the dilaton invariant, this is a perturbative symmetry
 in $\ti g$ perturbation theory.
We can consider orientifolding by
 $\ti \www$, which should make sense in $\ti g$ perturbation expansion.
It should be isomorphic to the usual orientifold construction, and lead to a
theory isomorphic to the usual type-I
theory, and in particular should have open strings and $SO(32)$ gauge symmetry.
The massless bosonic fields that are invariant under  $\ti \www$ are $g_{MN},
B^1_{MN}, \Phi$, which occurred in
the NS-NS sector of the (1,0) string for $g << 1$.
The Chan-Paton factors are now carried by the 9-branes that are the D-branes
for the (0,1) string; these are
precisely the (1,0) 9-branes that we have postulated.

The next question is that of how the \lq new' type I string is related to the
usual one. It is isomorphic to the usual
one, but it fits differently into the type IIB theory -- one is obtained by the
usual orientifold construction, the other by
doing an $SL(2,\Z)$ transformation and then orientifolding. The two type I
theories are physically indistinguishable;   both
are theories of non-orientible closed strings coupled to  open strings with
$SO(32)$ Chan-Paton factors and both  have $N=1$
supergravity coupled to $SO(32)$ super-Yang-Mills as the low energy effective
field theory.
On modding out by $\www$, the (0,1)  string, 5-brane and
9-brane of the IIB theory survive as type I  D-branes, while the (1,0) string
which is fundamental
for $g <<1$ leads
to the fundamental type I strings which can end on the D-branes. The  strong
coupling limit of this type I theory is
conjectured to be the $SO(32)$ heterotic theory [\Witten], with the (0,1)
string becoming the fundamental heterotic string
at strong coupling [\Oopen,\dab,\PW].
On the other hand,
on modding out by $\ti \www$ for $g >> 1$, the (1,0)  string, 5-brane and
9-brane of the IIB theory survive as the new type I  D-branes, while the (0,1)
string which is fundamental in the IIB theory
for $\ti g <<1$ leads
to the fundamental strings of this new type I  theory which can end on the
(1,0)  string, 5-brane and
9-brane.

To proceed further, we shall suppose that the perturbative parity reversal
symmetry $\www$ extends
to a
$\Z _2 $
 symmetry
of the full non-perturbative type IIB theory, which we will again denote as
$\www$. We know how this acts on perturbative states and hence on the effective
low-energy supergravity theory --
$g_{MN}, B^2_{MN}, \Phi$ are even and
$D_{MNPQ},  B^1_{MN}, \chi$ are odd under $\www$ -- and this tells us how it
acts on BPS states:
the  (0,1)  string, 5-brane and
9-brane are invariant while the the (1,0)  string, 5-brane and
9-brane
together with the 7-brane and 3-brane are odd under $\www$.
If there is such a non-perturbative symmetry $\www$ it should in particular be
a symmetry for all values of the coupling $g$.
Then $\ti \www$ should also extend to a non-perturbative symmetry and should be
related to $\www$ by
 $$ \ti \www = S\www S^{-1}
\eqn\tiw$$
where $S$ is the $SL(2,\Z)$ transformation \soi.
Modding the non-perturbative IIB theory by the
$\Z _2 $ symmetry
$\www$ (or $\ti \www$) should then make sense for any value of the coupling $g$
and would give   a theory with $N=1$ supersymmetry in $D=10$ and $SO(32)$ gauge
symmetry.
Modding out by $\www$ requires a background with 32  $(0,1)$ 9-branes (since it
does for weak
coupling, it must for strong coupling also) while modding out by $\ti \www$
requires 32 $(1,0)$
9-branes. As modding out by $\www$ at weak coupling ($g << 1$) gives the type I
string, and the
strong coupling limit of this is the $SO(32)$  heterotic string, this implies
that modding the
strongly coupled IIB string   (with 32 $(0,1)$ 9-branes)  out by $\www$ should
give the heterotic
string with coupling constant $\ti g=1/g$. Similarly, modding out the weakly
coupled type IIB
string (with 32 $(1,0)$ 9-branes) by $\ti \www$ should give the weakly coupled
heterotic string
with coupling constant $g$.  The NS-NS sector of the type IIB string  $g_{MN},
B^1_{MN}, \Phi$
gives the $N=1$ supergravity fields of the heterotic string and in particular
the type IIB dilaton
is identified with the heterotic dilaton.

The symmetry $\ti
\www$ should be a perturbative symmetry of the weakly coupled IIB string.
The weakly coupled type IIB string also has a perturbative symmetry
$(-1)^{F_L}$ where $F_L$ is the
left-handed fermion number of the conventional NSR formulation of the type IIB
theory. It also
leaves
$g_{MN}, B^1_{MN}, \Phi$ invariant -- the same sector preserved by $\ti \www$.
This implies that $(-1)^{F_L}$ and  $\ti \www$ act in the same way on BPS
states -- D-branes are odd and
(1,0) strings and 5-branes are even under both.
This suggests that    $(-1)^{F_L}$
is the same symmetry as $\ti \www$, and this could be taken as the definition
of the extension of
$(-1)^{F_L}$  to a symmetry of the full non-perturbative theory.
Indeed, in [\Se] it was argued
that
$(-1)^{F_L}=S\www S^{-1}$ by considering the action on massless fields, while
here
$\ti \www = S\www S^{-1}$ by definition.
Similarly, extrapolating $\www$ to strong coupling should give $(-1)^{F'_L}$
where $F'_L$ is the
left-handed fermion number of the   NSR formulation of the dual type IIB
theory, acting on fermions
moving on the (0,1) string world-sheet.

It has been suggested  (see e.g.  [\Se]) that modding out the type IIB theory
by $(-1)^{F_L}$
should give the type IIA theory  by considering the action on perturbative
states, while we have
argued that modding out by $\ti \www$ should give a heterotic/type I theory.
If $(-1)^{F_L}$ and  $\ti \www$ are in fact the same symmetry and if it is
indeed the case  that
modding   the type IIB theory by $(-1)^{F_L}$
can  give  the type IIA theory, then this would mean  that
there are two different ways of modding the IIB theory by the same symmetry, in
which different \lq twisted sectors' are added
to obtain a consistent theory. In both cases the untwisted or invariant sector
includes the
$g$-perturbative states from the product of the left-handed NS sector with
the right-handed R and NS sectors.
This theory is inconsistent as it stands, but a consistent theory can be
obtained by adding a \lq twisted sector'.
One way of doing this is by introducing   a left-handed R sector of the
opposite chirality, to obtain a type IIA theory, while
another is to introduce 32 (1,0) 9-branes and a heterotic sector which becomes
the open string sector of a type I theory as $g
\to
\infty$. If this is correct, it would also mean that on taking  an orientifold
of the type IIB theory with $\www$,
the resulting theory can be fixed up   either in the usual way to obtain the
type I string, or by adding a IIA twisted sector
so that the theory becomes the weakly coupled IIA theory as the IIB coupling
becomes large, so that the IIA and IIB couplings
are inversely related. As the  IIA theory becomes M-theory at strong coupling
[\Witten], this would mean that the twisted
sector added to the orientifolded type IIB theory at weak coupling should be
11-dimensional.
(It is of course conceivable that
    $(-1)^{F_L}$ and $ \ti
\www$ are in fact two different $\Z _2$ symmetries, which have the same action
on the massless sector, but differ in their
actions on the full theory; modding out by one gives the type IIA theory and
modding out by the other gives a   type I
theory. This would mean that $X=(-1)^{F_L} . \ti
\www$ should be some as yet unknown symmetry that preserves the massless and
BPS sectors but acts non-trivially on the full
theory, so that $ \ti
\www= (-1)^{F_L}.X $.)

This can be generalised to consider orientifolding the $(p,q)$ string instead
of the $(1,0)$ or
 $(0,1)$ string.
Acting with an $SL(2,\Z)$ transformation $\lll$
gives the discrete symmetry
$$\www _\lll = \lll \www \lll^{-1}
\eqn\abc$$
If $\lll$ takes a $(1,0)$ string to a $(p,q)$ string and takes the coupling
constant $g$ to $g_\lll$, then
perturbation theory in $g_\lll$ gives a type IIB theory in which the $(p,q)$
string is now the fundamental string
and $\www_\lll$ acts as parity reversal on the $(p,q)$ string world-sheet.
Modding out by  $\www_\lll$ should again give a type I string (for weak
$g_\lll$) and a heterotic string (for strong $g_\lll$);
each choice of $\lll$ gives a physically equivalent theory, but \lq embedded'
differently in the type IIB string.

\chapter{The IIA Theory and M-Theory}

We have argued that there should be an extra 9-brane in the type IIB theory.
Dimensionally reducing on a circle to 9 dimensions, the
9-brane yields a new 8-brane in the $D=9$ type II theory, in addition to the
D-8-brane that comes
from the RR 9-brane in the IIB theory. The same $D=9$ type II theory can be
obtained by reducing the
T-dual type IIA theory, so that the new 8-brane should also have a type IIA
origin. It could have
come from   either a new 8-brane or a new 9-brane in the $D=10 $ type IIA
theory; however, a new
type IIA 8-brane would give both an 8-brane and a 7-brane in $D=9$, and there
is no evidence for an
extra 7-brane from the type IIB side. Thus the type IIA theory should also have
a  new 9-brane,
which again is not in the RR sector and is not a D-brane. The type IIA theory
at finite coupling is
M-theory compactified on a circle, so that the new type IIA 9-brane should have
an M-theory origin.
The simplest way in which this could happen would be if    M-theory had either
a   9-brane or a   10-brane. A
10-brane would give just a 9-brane in $D=10$, while a 9-brane would give a
9-brane and an 8-brane. We have seen in
section 2 that the $D=11$ superalgebra \el\ has a 9-brane charge that reduces
to an 8-brane charge and
a 9-brane charge in $D=10$, so we conclude that M-theory could have  a 9-brane
that gives rise to the new
9-brane and the 8-D-brane of type IIA on reducing on a circle.

Assuming the existence of an M-9-brane,  M-theory then  has $p$-branes for
$p=2,5,6,9$, and reducing to the $D=10$ IIA theory,
the M-membrane gives  branes with $p=1,2$, the M-5-brane gives  branes with
$p=4,5$, the M-6-brane gives  branes
with $p=5,6$ and the M-9-brane gives  branes with $p=8,9$.  In addition, the
11-momentum $P_M$ gives
a 0-brane charge and a 10-momentum on reduction. If the $D$-momentum $P_M$ were
treated in the same
way as the other charges, it would be split into the spatial momentum $P_i$
which is analogous to a
1-brane charge, and the energy $P_0$, which could be dualised to a $D-1$ brane
charge
$\hat P_{i_1...i_{D-1}}$. The fact that states carrying momentum in
compactified dimensions are
U-dual to states obtained by wrapping branes round the  compactified dimensions
[\HT] means that for
some purposes states carrying the gravitational charge $P_M$ should be treated
on the same footing
as states carrying brane charges.

If the $D=11$ momentum $P_M$ is formally thought of as corresponding to a
1-brane charge $P_i$ and a 10-brane charge $\hat
P_{i_1...i_{D-1}}$ (dual to $P_0$), then the M-theory spectrum has branes with
$p=1,2; 5,6; 9,10$. It is
   suggestive that
these occur in
 three pairs
$\{ p-1,p\}$ for $p=2,6,10$. This would have a natural interpretation if these
charges arose from a
12-dimensional theory with a 2-brane, a 6-brane and a 10-brane.
Twelve-dimensional theories have
been proposed for a number of different reasons [\HStr,\Twe,\Tse].
If the 12-dimensional space has signature $(11,1)$, then this reduction is
straightforward.
It has been suggested [\Twe]
that the superalgebra in 10+2 dimensions [\TwAl]
$$
\{Q_\alpha,Q_\beta\} =
\big(\Gamma^{MN}C\big)_{\alpha\beta}\, Z_{MN} +
\big(\Gamma^{MNPQRS}C\big)_{\alpha\beta}\, Z^+_{MNPQRS}\ .
\eqn\eltw$$
(with $Q_\aa$ 32-component Majorana-Weyl spinors, and $Z^+_{MNPQRS}$ a
self-dual 6-form) should play
a role in 12 dimensions.  On dimensional reduction with respect to one of the
time-like directions,
the $D=11$ algebra \el\ emerges, with the momentum $P_M$ arising from the
$D=12$ 2-form charge
$Z_{MN}$. Choosing one of the two times $x^0$ as the \lq canonical time', the
remaining coordinates
are the other time coordinate $x^{\hat 0}$ and the spatial coordinates $x^i$
{}$(i=1,...,10)$, which
can be combined into 11-dimensional coordinates $x^\mu$, with $\mu=\hat 0,i$.
The  2 charges
$Z_{MN},Z^+_{MNPQRS}$ give rise to a 2-brane charge $Z_{\mm\nn}$ and a 6-brane
charge
$Z_{\mm_1...\mm_6}$, while
$Z_{0\mm}$ is dualised to give  a 10-brane charge $\hat Z_{\mm_1...\mm_{10}}$;
the dual 6-brane
charge $\hat Z _{\mm_1...\mm_6}$ is equal to
$Z_{\mm_1...\mm_6}$ because of the self-duality of $Z^+_{MNPQRS}$.  Each of
these $p$-brane charges
with $p=2,6,10$ splits into  one including an $x^{\hat 0}$ component and one
without, giving the
charge for a brane with world-volume of signature $(p,1)$ and one for a brane
with world-volume of
signature $(p-1,2)$.
On dimensionally reducing in the $x^{\hat 0}$ direction, one obtains in 11
dimensions a $p$-brane and a
$p-1$ brane, both with
conventional Lorentzian signature.

 \vskip 1cm

\begintable
 12-Brane | World-Volume Signature | Charge | M-brane \elt
 2-brane | (2,1) | $Z_{ij}$| 2-brane \elt
 (1,1)-brane | (1,2) | $Z_{i\hat 0}$| pp-wave \elt
 6-brane | (6,1) | $Z_{ijklmn}$| KK monopole \elt
 (5,1)-brane | (5,2) | $Z_{ijklm\hat 0}$| 5-brane \elt
10-brane | (10,1) | $Z_{i_1...i_{10}}$| Energy $P_0$ \elt
 (9,1)-brane | (9,2) | $Z_{i_1...i_9\hat 0}$| 9-brane
\endtable

\centerline{{\bf Table 1} Suggested branes in 12 dimensions and reduction to 11
dimensions.}

The conventional type IIA supergravity has no
8-brane solution, but it was shown in [\Ebr] that Romans'   generalisation of
the type IIA theory involving a mass $m$ [\Rom]
does have  such a solution. The 8-brane couples to a non-dynamical 9-form
potential $A_9$ with field strength $F_{10}$. The
massive type IIA action was rewritten in [\Ebr] in a form which included the
term
$$\int M F_{10}
\eqn\MF$$
where $M$ is a scalar field. As $A_9$ only occurs in the action through this
term, it is a Lagrange multiplier imposing the
constraint $M=constant$ and the constant  value of $M$ is the mass $m$ of the
Romans theory. If $m=0$, the standard type IIA
supergravity is recovered. This 8-brane should arise from
the double dimensional reduction of the  9-brane of M-theory, and the term
\MF\ should also have an 11-dimensional origin.
Indeed, it was argued in [\Ebr] that the term \MF\ necessarily arises in the
IIA string theory,
which implies that a similar term must arise in M-theory, as the M-theory
compactified on a circle is  the same thing as the type IIA
string at finite coupling.
This suggests that the low-energy effective field theory of M-theory
should include  a term
$$\int M F_{11}
\eqn\abefr$$
where $M$ is a scalar and $F_{11}=dA_{10}$ where $A_{10}$ is a 10-form
potential which couples to the M-theory 9-brane.
On reduction to $D=10$, the potential $A_{10}$ gives $A_9$ which couples to the
IIA 8-brane, and a 10-form $A_{10}$
which couples to the 9-brane. However, it has been shown
subject to certain assumptions that there is
no conventional   extension of
$D=11$  supergravity which includes  a term of the form \abefr\ [\Des], so that
any modification of
the supergravity theory including the term \abefr\
must be of an unusual form.

\chapter{World-Volume Theories and Collective Coordinates}

The zero-mode dynamics of  a $p$-brane are described by a  supersymmetric
 field theory on the
$p+1$ dimensional world-volume arising from gauge-fixing  a $\kk$-symmetric
supercovariant action.
This world-volume theory has 8 fermionic and 8 bosonic degrees of freedom for
branes that break half
the supersymmetry. The bosonic degrees of freedom for a $p$-brane usually
include $D-p-1$ scalar
fields, corresponding to the collective  coordinates for the position of the
$p$-brane, and if $D-p-1<8$, there are in addition $9+p-D$ bosonic degrees of
freedom which are
typically collective coordinates for anti-symmetric tensor gauge field degrees
of freedom.
The $D=10$ fundamental string is described by   8 world-sheet scalars and 8
world-sheet spinors, while the
  $D=11$ membrane is described by  8 world-volume scalars and 8
3-dimensional spinors;
the scalars transform as a vector and the fermions as a spinor under the
transverse $SO(8)$.
 A
D-brane is described by  a   vector multiplet in its $p+1$ dimensional
world-volume
with $9-p$ scalars and a vector
field, giving $(9-p)+(p-1)=8$ bosonic physical  degrees
 of freedom. The 5-brane of M-theory is described by a 6-dimensional self-dual
vector multiplet
whose bosonic degrees of freedom  consist of 5 scalars and a 2-form gauge field
with self-dual
field strength.

The description of the dynamics of the  membrane in $D=11$ requires 8 scalars,
while
the $D=10$ membrane of the type IIA theory should have one less scalar
collective coordinate; this
is indeed the case, and its effective world-volume theory  has 7 scalars and a
vector. In the
three-dimensional world-volume, a scalar is dual to a vector, and the two
membrane effective
actions are related by a world-volume duality transformation.
Similarly, the type IIA 5-brane
should have 4 scalars on its world-volume, while
 the M-theory 5-brane
from which it  is obtained by double dimensional reduction has 5 scalars.
 This suggests that the
IIA 5-brane should be described by the 6-dimensional multiplet obtained by
taking the self-dual
antisymmetric tensor multiplet and dualising one of the scalars to obtain a
4-form. The scalar to be dualised
is the one corresponding to translations in the 11-th dimension, which is taken
to be an isometry in the dimensional
reduction, so that the scalar only appears through its derivative and can be
dualised.
 Thus the bosonic sector of the M-theory 5-brane effective theory has 5 scalars
and a
self-dual 2-form gauge field, the IIA 5-brane has 4 scalars, a self-dual
2-form gauge field and a 4-form gauge field; the two multiplets are on-shell
equivalent.
 In the IIB
theory, the D-5-brane is described by a vector multiplet with 4 scalars and a
vector. This can be
dualised to a multiplet with 4 scalars and a  3-form gauge field, and this is a
natural candidate for
the world-volume theory of the NS 5-brane of the type IIB theory.

In [\Tse,\Dua], explicit duality transformations were considered for
D-brane actions, and for the cases considered
the results were consistent with the view that the transition from weak to
strong coupling is
accompanied by a world-volume duality transformation. However, the
determination of the zero modes
from the supergravity field equations,
as in [\CHS], does not distinguish between dual forms of the world-volume
collective coordinate
supermultiplet;
  the choice between dual forms of the multiplets
seems to be a matter of convenience rather than of principle. For example, for
the membrane the formulation with eight
 scalars makes the (transverse part of the) 11-dimensional Lorentz invariance
manifest, while
it is more convenient to write the non-abelian generalisation arising when a
number of membranes are coincident in terms of
the vector multiplet, as the corresponding action written in terms of the
scalars would be non-local.
In what follows, we shall seek dual forms of the multiplet in which the number
of scalars is the number of translational
zero-modes, and   different branes have different multiplets.

We now turn to the question of the KK monopoles. We have argued that they can
be regarded as $D-5$
branes, which should imply that they have 4 scalar degrees of freedom (since a
$p$-brane should
have
$D-p-1$ scalar fields).
However, this cannot be correct, as this number of scalars cannot be fitted
into supermultiplets
of half-maximal supersymmetry in general. For example, for $D=11$ we would need
a 7-dimensional
$N=2$ supermultiplet with 4 scalars, of which none are known. In 7 dimensions,
an $N=2$ vector multiplet has
three scalars and a vector, and the number of scalars can be decreased by
dualising, but cannot be
increased. The resolution comes from considering the gravitational instanton
moduli space.
The space-time in question is $\R^{p,1}\times N_4$ where $N_4$ is a
gravitational multi-instanton space.
For a single instanton in four dimensions, one would expect four collective
coordinates
corresponding to the position of the instanton. However, this is not the case
for the Taub-NUT
metric, which has a 3-dimensional moduli space; translation in the  direction
$\pa/\pa y$ gives an equivalent metric and so
does not count as a deformation as that direction is Killing [\Rub]. More
generally,
the multi-centre metrics \remet,\vis,\gist, have only $3s$ moduli
(corresponding to the $s$ positions $x_i^\mm$ in
$\R^3$) [\Rub]. This would be consistent with
the effective world-volume dynamics of the KK monopole being described by a
7-dimensional vector
multiplet,
which has 3 scalars, or by the multiplet given by dualising the vector to give
3 scalars and a
4-form gauge field;    these are the only possibilities as no other matter
multiplets have 3
scalars. It is remarkable that supersymmetry would have led us to the
conclusion that the moduli space
dimension should be a
   multiple of 3.

Double dimensional reduction to the  KK monopole 5-brane in 10 dimensions
gives the
dimensional reduction of the vector multiplet to $D=6$, and should contain 3
scalars  again, corresponding
to the  instanton moduli. This leads to the multiplet with 3 scalars, one
3-form and one 4-form
gauge field, coming from the reduction of the multiplet with 3 scalars and a
4-form in 7 dimensions.
This suggests that a convenient description is in terms of a   multiplet with 3
scalars and a 4-form
in 7 dimensions or the multiplet obtained by straightforward dimensional
reduction of this to 7-d
dimensions for the KK monopole in $D-d $ dimensions.

Simple dimensional reduction of the KK monopole (taking a periodic array)
 gives the type IIA 6-brane in $D=10$.
 The    7-dimensional world-volume description is obtained by dualising the
 4-form to a vector to give the
 $D=7$ vector multiplet with 3 scalars and a vector. The picture that seems to
be emerging is that
double dimensional reduction of an M-theory $p$-brane for $p=2,5,6$ gives a
$p-1$ brane of the IIA
theory with the world-volume theory obtained by dimensional reduction of the
M-brane world-volume
theory from $p+1$ to $p$ dimensions. On the other hand, simple dimensional
reduction (with a
periodic array) to give a IIA $p$-brane gives   equivalent world-volume
theories for the IIA and
M-theory $p$-branes, but related by a world-volume duality transformation.
If this pattern persists for the M-theory 9-brane, its world-volume theory must
be a 10-dimensional
supersymmetric theory whose dimensional reduction to 9 dimensions
gives the 8-brane world-volume theory which is given by a vector multiplet with
a vector and a
scalar. This fixes the M-theory 9-brane to be described by a 10-dimensional
vector multiplet.
One would have expected a scalar representing the position of the 9-brane, but
perhaps this is
missing for  reasons similar to those for the absence of the expected fourth
modulus for the
Kaluza-Klein monopole. Then the world-volume theory for the IIA 9-brane should
be the multiplet
related to this by a world-volume duality transformation, with a 7-form gauge
potential instead of a
vector.

Consider now the IIB theory. As the solitonic (NS) 5-brane is related to the
Kaluza-Klein monopole
by T-duality (with respect to the isometry generated by $\pa /\pa y$ for the
metric \remet),
the IIA theory in a 5-brane background is equivalent to the IIB theory in a KK
monopole background,
and the IIB theory in a solitonic  5-brane background is equivalent to the IIA
theory in a KK monopole
background. This fixes the world-volume theory of the IIB KK monopole to be the
  six
dimensional (2,0) supermultiplet with a 2-form $A_2^+$ with self-dual field
strength and 5 scalars,
or a dual version of this. As there are 3 translational zero modes, it is
natural to seek a dual
form with 3 scalars; if it is possible to dualise 2 scalars, this would give a
 multiplet
with $A_2^+$, 3 scalars and two 4-form potentials $A_4$. (Note that the five
scalars could have interpretation as
translational  zero modes of 7-brane in 12 dimensions which reduces to this
5-brane in 10 dimensions.)
It also fixes the six-dimensional world-volume multiplet for the type IIB NS-NS
5-brane to be the
(1,1) vector supermultiplet with 4 scalars and a vector,
or a dual version of this. There should be 4 translational zero modes and so 4
scalars, but the
vector could be dualised to give a multiplet with   4 scalars  and a 3-form.

Finally, we consider the pp-wave solutions. As there is a sense in which they
can be regarded as
1-branes and are T-dual to fundamental strings, we can ask what their
world-sheet effective dynamics should be.
The pp-wave solution of M-theory [\Hwave] gives the 0-brane of the IIA theory
on double dimensional
reduction (i.e. on reducing the configuration in which the wave travels in the
compact 11th
dimension) [\PKT] and as the 0-brane world-line theory is a one-dimensional
super-vector multiplet
with a vector gauge field and 9 scalars, the pp-wave world-sheet theory which
reduces to this must
be a 2-dimensional theory with 8 scalars and a vector. If the pp-wave is moving
in a compact
dimension, then it is related by T-duality to a fundamental string wrapped
around the compact
dimension [\PPdual]. Thus the IIA
theory in such a pp-wave background is  T-dual to the IIB theory in a
fundamental string background, i.e. with a fundamental string wrapped around
the internal
dimension [\BHO]. This fundamental string is in turn dual to a D-string. The
effective world-sheet
dynamics is then described by a IIB Green-Schwarz string with 8+8 degrees of
freedom in static
gauge, or equivalently by a Born-Infeld action
which is equivalent to this via a world-sheet duality transformation [\Dua].
Similarly, the IIB pp-wave is T-dual to the IIA fundamental string [\BHO], and
is described by a IIA
Green-Schwarz superstring action.

These world-volume dynamics are summarised in the following tables. The bosonic
fields in the supermultiplet  are listed, with
$A_n$ denoting an $n$-form gauge potential, $A_2^+$ denoting a 2-form with
self-dual field strength and $m\times \phi$
represents $m$ scalars. The multiplets for the solitonic 5-branes have also
been proposed in
[\Berg].

 \vskip 1cm

\begintable
 M-Brane | World-Volume Multiplet | Suggested Dual Form \elt
 pp-Wave | $A_1,~8 \times \phi$ | $A_1,~8 \times \phi$ \elt
 2-Brane  | $8\times \phi$ | $8\times \phi$ \elt
 5-Brane | $A^+_2, ~ 5\times \phi$ | $A^+_2, ~ 5\times \phi$ \elt
 6-Brane (KK Monopole) | $A_1,~3 \times \phi$ | $A_4,~3 \times \phi$ \elt
 9-Brane | $A_1$ | $A_1$
\endtable

\centerline{{\bf Table 2} World-Volume Dynamics of M-Branes.}

 \vskip .5cm

\begintable
 IIA-Brane | World-Volume Multiplet | Suggested Dual Form \elt
 $\hbox{Dirichlet p-Brane} \atop \hbox{p=0,2,4,6,8}$ | $A_1,~(9-p) \times \phi$
| $A_1,~(9-p) \times \phi$ \elt
 1-Brane (Fundamental) | $8\times \phi$ | $8\times \phi$ \elt
 pp-Wave | $A_1, ~ 8\times \phi$ | $A_1, ~ 8\times \phi$ \elt
 5-Brane (Solitonic) | $A^+_2, ~ 5\times \phi$  | $A^+_2, ~ A_4,~ 4\times \phi$
\elt
 5-Brane (KK Monopole) | $A_1,~ 4\times \phi$ | $A_4,~A_3,~ 3\times \phi$  \elt
 9-Brane (Extra) | $A_1$ | $A_7$
\endtable

\centerline{{\bf Table 3} World-Volume Dynamics of Type IIA Branes.}

 \vskip .5cm

\begintable
 IIB-Brane | World-Volume Multiplet | Suggested Dual Form \elt
 $\hbox{Dirichlet p-Brane} \atop \hbox{p=1,3,5,7,9}$ | $A_1,~(9-p) \times \phi$
| $A_1,~(9-p) \times \phi$ \elt
 1-Brane (Fundamental) | $8\times \phi$ | $8\times \phi$ \elt
 pp-Wave | $ 8\times \phi$ | $ 8\times \phi$ \elt
 5-Brane (Solitonic) | $A_1,~ 4\times \phi$ | $A_3,~ 4\times \phi$ \elt
 5-Brane (KK Monopole) | $A^+_2, ~ 5\times \phi$ | $A^+_2, ~ 3\times
\phi,~2\times A_4$ \elt
 9-Brane (Extra) | $A_1$ | $A_7$
\endtable

\centerline{{\bf Table 4} World-Volume Dynamics of Type IIB Branes.}

 \vskip .5cm

In the last section, the possibility that the branes of M-theory could have a
12-dimensional origin was discussed.
If it is indeed the case that there is a 2-brane, 6-brane and 10-brane in 12
dimensions which reduce to the M-branes, it is
interesting to ask whether consistent world-volume dynamics could be attributed
to them which gives the correct dynamics on
dimensional reduction. The 2-brane in 12 dimensions should give rise to the
2-brane and pp-wave in 11 dimensions, both of
which have 8 scalars, and so should itself have 8 scalars; this is one less
than would have been expected for a standard
2-brane in 12 dimensions. The 6-brane is more problematic. On reduction to
M-theory, it should give both the 5 brane with a
6-dimensional self-dual tensor multiplet and the 6 brane with a 7-dimensional
vector multiplet. If the $D=12$ signature is
(11,1), the 6-brane world-volume has signature (6,1) and there is no
7-dimensional multiplet that is equivalent to the vector
multiplet and reduces to the tensor multiplet. If the
$D=12$ signature is (10,2), the M-theory 6-brane arises from the
$D=12$ 6-brane with world-volume   signature (6,1), and this should then have a
world-volume vector multiplet.
The M-theory 5-brane arises from the
$D=12$ 6-brane with world-volume   signature (5,2),  and this should be
described by a  multiplet
in (5,2) dimensions that reduces to the  self-dual tensor multiplet in (5,1)
dimensions. In particular, chiral
fermions and a self-dual tensor must emerge on dimensional reduction, and this
could not happen if
the (5,2) theory was a conventional local theory with $O(5,2)$ Lorentz
invariance and with a
standard dimensional reduction. Thus it appears that either the $D=12$ branes
cannot have
covariant dynamics -- it might involve a particular vector, such as a fixed
null vector   -- or the
reduction to
$D=11$ must be non-standard.

\chapter{ALE Branes, ALF Branes and Symmetry Enhancement}

When parallel D-branes of the same type approach one another, the world-volume
gauge symmetry is
enhanced to a non-abelian group with extra massless states arising from
fundamental strings joining
the D-branes. The symmetry enhancement should be true for all values of the
coupling, although the
interpretation can change. The type IIA string becomes 11-dimensional M-theory
at strong coupling,
and D-branes for
$p=0,2,4,5,6,8$ become,
 as we have seen, pp-waves, 2-branes, 5-branes, KK monopoles and 9-branes
respectively, which can be thought of as M-branes with $p=1,2,5,6,9$. The
configuration of two
$p$-D-branes joined by fundamental strings, for which the world-volume theory
has a $U(2)$ gauge
symmetry when the branes coincide, becomes two M-branes joined by a 2-brane in
$D=11$. Indeed, it
was shown in [\BR] (and references therein) that an M-theory 2-brane can
intersect a pp-wave,
2-brane, 5-brane or KK monopole in such a way that breaks $1/2$ the
supersymmetry and which reduces
to a fundamental string intersecting a D-brane with $p=0,2,4,6$ respectively.
Then two
parallel M-branes of the same type (pp-waves, 2-branes, 5-branes, KK monopoles
or 9-branes)
give enhanced symmetry when they come together,
giving  a non-abelian gauge theory or, in the case of 5-branes, a theory of
tensionless self-dual strings [\Strom].
 As a result, in the IIA theory two fundamental
strings, two pp-waves, two 5-branes or KK monopoles coming together gives
enhanced symmetry.
For the type IIB theory, enhanced symmetry also results from bringing together
 two fundamental strings, two solitonic 5-branes or two (1,0) 9-branes,
  as these are
  related by $SL(2,\Z)$ duality to D-branes, and from
 bringing together two KK monopoles or two pp-waves,
 as these are
  related by
 T-duality to type IIA 5-branes and fundamental strings, respectively. We now
examine this in more
detail for the case of KK monopoles.\foot{Some of the following was also
considered in [\PKTlecs].}

The Gibbons-Hawking gravitational multi-instanton metric [\Haw,\GHaw] is
$$ds^2=V(d y + A_\mu d x^ \mu)^2 +V^{-1}\dd_{\mu \nu}d x^\mm
d x^\nn
\eqn\remeta$$
where
$$ V^{-1}=\epsilon
+\sum _{i=1}^s {2n  \over \vert x^\mm-x_i^\mm\vert },
\qquad
F_{\mm\nn}=\ee
_{\mm\nn\rr}\nabla^\rr V^{-1}
\eqn\abc$$
If  $\epsilon=1$, this is the multi-Taub-NUT space considered earlier with ALF
(asymptotically locally  flat) boundary
conditions. If $\epsilon=0$,  $n$ can be scaled to $1$ and the space is a
multi-centre generalisation of the
Eguchi-Hanson metric with  ALE (asymptotically locally  Euclidean) boundary
conditions. The case $\epsilon=0, s=1$ is flat space, while
$\epsilon=0, s=2$ gives the Eguchi-Hanson instanton.
The space $\R^{D-5,1} \times N_4$ where $N_4$ is an  ALF or ALE instanton is a
solution of M-theory $(D=11)$ or of string
theory ($D=10$) as the gravitational instanton is hyperkahler [\AGH].

The multi-Eguchi-Hanson space behaves as $\R^4/ \Z_s$ at large distances, and
can be used to resolve
an $A_s$ singularity of an orbifold limit of $K_3$, or blow up the orbifold
singularity of $\R^4/ \Z_s$. For example, one
orbifold limit of
$K_3$ is
$T^4/\Z_2$ and each of the 16 orbifold singularities can be repaired by gluing
in an Eguchi-Hanson
metric. In the limit $\vert x_1-x_2\vert  \to 0$, the Eguchi-Hanson space
becomes $\R^4/ \Z_2$ with
an orbifold singularity at $x_1$.  While the multi-Taub-NUT solution can be
thought of as a
solution with a number of parallel $D-5$ G-branes, the multi-Eguchi-Hanson
space can be viewed as a
number of parallel $D-5$ G-branes
 in  a transverse space which is an orbifold; as a result, not all the branes
are independent as
some are \lq mirror images' of others. There is one instanton corresponding to
each pair
$\{x_i,x_j\}$.

As the $y$ coordinate is periodic, each line segment in the $\R^3$
parameterised by $x^\mu$ is
associated with a cylinder in
$N_4$, unless the line segment passes through one of the points $x_i^\mu$ at
which the size of the
$y$-circle shrinks to zero. In particular, a line segment joining $x_i^\mu$ and
 $x_j^\mu$
corresponds to a 2-sphere and the set of all such 2-spheres corresponding to
all pairs $\{x_i,x_j\}$
forms a basis for the second homology. As one approaches a point in moduli
space at which $\vert
x_i-x_j\vert
\to 0$, the area of the corresponding 2-sphere shrinks to zero and this is
associated with symmetry
enhancement.

For type IIA string theory compactified on $K_3$, the equivalence with
heterotic string theory
implies the existence of special points in the $K_3$ moduli space at which the
gauge symmetry is
enhanced [\HT]. Indeed, it follows from supersymmetry that the mass of certain
BPS states tends to
zero at these special points [\HUSC,\HTE] to give the extra massless vector
multiplets. At these
points the area of certain homology 2-cycles shrinks to zero to give an
orbifold limit of $K_3$ and
the BPS states in 6 dimensions arising from 2-branes wrapped around the
shrinking 2-cycles become
massless [\Witten,\HUSC,\HTE].  The behaviour of the theory as a particular
set of 2-cycles shrinks
can be studied by looking at the theory on the   space $\R^{D-5,1} \times N_4$
in which the $K_3$ is
replaced by the appropriate multi-Eguchi-Hanson space [\BSV]; this gives a good
approximation when
the radius of the  2-sphere is small compared to the size of the $K_3$. The
enhanced gauge symmetry
is then associated with parallel G-branes becoming coincident, so that homology
2-spheres shrink to
zero size and the  membranes wrapped around these
 give rise to massless states on the G-brane so that the G-brane world-volume
theory becomes a
non-abelian gauge theory. (The   world-volume theory for each G-brane is a
$D=6$ vector multiplet,
as for the ALF case.) This is straightforward to check using duality; a
T-duality relates the
solution to a multi-centre solitonic 5-brane solution of type IIB theory (in
$\R^4/ \Z_s$) and $SL(2,\Z)$ duality relates this to a     configuration of
parallel 5-D-branes, and
there is symmetry enhancement as these become coincident due to strings joining
the 5-D-branes
becoming massless [\BSV]. In the case of the Eguchi-Hanson space ($s=2$), the
instanton shrinks to
zero size as the 2-sphere shrinks, so that the symmetry enhancement is a
gravitational analogue of
the symmetry enhancement as certain Yang-Mills instantons shrink to zero size
[\Winstanton].

There is a similar symmetry enhancement as KK monopole G-branes become
coincident. This is to be expected from the fact that
 the 6-D-brane of type IIA arises from the KK monopole of M-theory [\PKT]; the
symmetry enhancement of the
world-volume theory as the 6-D-branes become coincident implies the same
enhancement as  parallel KK monopole G-branes
of M-theory become coincident, and this in turn implies a similar effect for KK
monopoles of the type IIA theory.
For the 6-D-brane, the extra massless states come from strings joining the
approaching 6-D-branes, while for the KK monopole
$D-5$ G-branes, these come from membranes wrapping around the shrinking
2-cycles. The position of each
$D-5$ G-brane corresponds to a point $x_i$ in $\R^3$ and there is a  2-sphere
for each pair $\{x_i,x_j\}$.
Thus when two parallel G-6-branes of M-theory approach one another, the $D=7
{}~U(1)$ vector supermultiplet  on each brane
combines with the BPS state from membranes wrapping the 2-cycle joining them to
give a $U(2)$ super-Yang-Mills   theory on
the common 7-dimensional world-volume. This enhancement can be understood in
the IIA case from duality using the argument
of [\BSV]; the two  parallel G-5-branes of the IIA theory are T-dual to two
parallel solitonic 5-branes of the
IIB theory, and these are $SL(2,\Z)$-dual to two
parallel 5-D-branes, with the usual $U(1)\times U(1) \to U(2)$ symmetry
enhancement on the world-volume.
The membrane wrapping the 2-cycle is replaced by the string joining the
D-branes.

Consider now the analogous situation in the IIB string on $\R^{5,1}\times K_3$
or $\R^{5,1}\times N_4$.
In this case, three branes wrapping the  2-cycles gives
rise to self-dual strings in 6 dimensions that become tensionless in the limit
in which the 2-cycle shrinks to zero area [\WitUSC].
In particular, the limit in which two parallel G-5-branes become coincident,
the world-volume theory changes from an abelian
self-dual anti-symmetric tensor multiplet to a non-abelian generalisation
resulting from self-dual strings becoming null.
This is related by T-duality to the case of 2 parallel 5-branes of the IIA
theory becoming coincident, which in turn becomes
the case of coincident 5-branes of M-theory at strong coupling; in this latter
case, the 3-brane  wrapping the 2-cycle have
been replaced by a  membrane ending on the two 5-branes [\Strom].

\vskip 0.5cm
\noindent {\it Acknowledgements:} I would like to thank Jerome Gauntlett,
Gary Gibbons, Michael Green, Ashoke Sen and  Paul Townsend  for valuable
discussions.

\refout

\bye